\documentclass[aps,pre,twocolumn,reprint,amssymb,amsmath,nobibnotes,nofootinbib,superscriptaddress,floatfix]{revtex4}
\usepackage[dvips]{graphicx}
\usepackage[hidelinks]{hyperref}
\usepackage{epsfig}
\usepackage{framed}
\usepackage{color}
\usepackage{xcolor}
\usepackage{textcomp}
\usepackage{setspace}
\usepackage{morefloats}
\usepackage[T1]{fontenc}
\usepackage[utf8]{inputenc}
\usepackage{cleveref}
\usepackage{ulem}
\usepackage{float}
\usepackage{bm}
\usepackage{comment}

\newcommand\blue[1]{\textcolor{blue}{#1}}

\begin{document}

\title{Stationary particle currents in sedimenting active matter wetting a wall}

\author{Matthieu Mangeat}
\affiliation{Center for Biophysics \& Department for Theoretical Physics, Saarland University, D-66123 Saarbr{\"u}cken, Germany.}

\author{Shauri Chakraborty}
\affiliation{Center for Biophysics \& Department for Theoretical Physics, Saarland University, D-66123 Saarbr{\"u}cken, Germany.}

\author{Adam Wysocki}
\affiliation{Center for Biophysics \& Department for Theoretical Physics, Saarland University, D-66123 Saarbr{\"u}cken, Germany.}

\author{Heiko Rieger}
\affiliation{Center for Biophysics \& Department for Theoretical Physics, Saarland University, D-66123 Saarbr{\"u}cken, Germany.}
\affiliation{INM – Leibniz Institute for New Materials, Campus D2 2, D-66123 Saarbr{\"u}cken, Germany.}

\begin{abstract}
Recently it was predicted, on the basis of a lattice gas model, that scalar active matter in a gravitational field would rise against gravity up a confining wall or inside a thin capillary – in spite of repulsive particle-wall interactions [Phys. Rev. Lett. {\bf 124}, 048001 (2020)]. In this paper we confirm this prediction with sedimenting active Brownian particles (ABPs) in a box numerically and elucidate the mechanism leading to the formation of a meniscus rising above the bulk of the sedimentation region. The height of the meniscus increases with the activity of the system, algebraically with the P\'eclet number. The formation of the meniscus is determined by a stationary circular particle current, a vortex, centered at the base of the meniscus, whose size and strength increase with the ABP activity. The origin of these vortices can be traced back to the confinement of the ABPs in a box: already the stationary state of ideal (non-interacting) ABPs without gravitation displays circular currents that arrange in a highly symmetric way in the eight octants of the box. Gravitation distorts this vortex configuration downward, leaving two major vortices at the two side walls, with a strong downward flow along the walls. Repulsive interactions between the ABPs change this situation only as soon as motility induced phase separation (MIPS) sets in and forms a dense, sedimented liquid region at the bottom, which pushes the center of the vortex upwards towards the liquid-gas interface. Self-propelled particles therefore represent an impressive realization of scalar active matter that forms stationary particle currents being able to perform visible work against gravity or any other external field, which we predict to be observable experimentally in active colloids under gravitation.
\end{abstract}

\maketitle


\section{Introduction}
\label{sec_intro}

Active matter is constituted of self-propelled particles, like motile microorganism, bacteria, cells, animals, or active colloids, which consume energy at small scales and convert it into a persistent motion, driving the system out-of-equilibrium~\cite{marchetti2013, elgeti2015, shaebani2020}. This energy is often redistributed as thermal agitation~\cite{tailleur2009}, but a deeper understanding of these active energy flows is needed to extract a valuable work~\cite{malgaretti2022}. Experiments and numerical evidence reported in recent literature show that active matter gives rise to nontrivial non-equilibrium steady states in presence of boundaries and obstacles, such as accumulation at walls~\cite{elgeti2013,lee2013,wagner2017,sepulveda2017}, ratchet effects~\cite{nikola2016}, and long-range depletion interactions~\cite{ro2021}. Sperm, {\it E. Coli} bacteria, or microalgae confined to an observation chamber have been found to strongly adhere to the walls~\cite{lauga2006, sartori2018, rothschild1963, ostapenko2018}.

Several minimal models and plausible mechanisms have been proposed so far to explain and predict this behavior. For example, the minimal model consists of spherically symmetric, active Brownian particles (ABPs) without alignment but with excluded volume interaction~\cite{romanczuk2012, solon2015}, belonging to the class of scalar active matter, like run-and-tumble particles~\cite{tailleur2008} and active lattice gas~\cite{thompson2011, kourbane2018}. These active particles behave like a passive fluid with particle-particle attractive interactions, since the collisions between them slow down the dynamics, and therefore effectively attract each other. Consequently, ABPs separate into low-speed (dense) and high-speed (dilute) phases, a phenomenon called {\itshape motility-induced phase separation} (MIPS)~\cite{cates2015}. Although being a dynamically arrested phase, the dense active phase seems not to be a glassy phase~\cite{paoluzzi2022}. This phenomenon is now well characterized in the context of ABPs~\cite{siebert2018}, also in presence of attraction between particles~\cite{redner2013}, or with polydisperse particles~\cite{kumar2021}. The mechanism leading to MIPS is also responsible for the wall accumulation of active particles~\cite{elgeti2013, lee2013}, due to the adhesion of ABPs on repulsive walls.

However, the effect of boundaries and steric interaction forces on active matter in the presence of an external force is not yet well understood. Experiments and Brownian dynamics simulations have shown that a system of dilute self-propelled particles -- chemically powered colloids -- sediment under an external gravitational field~\cite{palacci2010, enculescu2011, ginot2015, ginot2018}. The sedimentation length increases quadratically with the swimming velocity of the ABPs, and active particles can partially swim against the gravity~\cite{enculescu2011}. Exact steady-state solutions have been also derived in the context of 2d and 3d ideal active sedimentation~\cite{hermann2018, vachier2019}. However, it is not obvious how the combination of an external gravitational field and wall interactions might affect the steady state of a system of interacting active particles.

Recently, in the context of understanding the phenomenon of capillary action and spontaneous imbibition of liquids in porous media, a minimal active lattice gas model consisting of self-propelled hard-core particles in an external gravitational field had been introduced~\cite{wysocki2020}. By inserting a thin capillary tube into the bulk-sedimented phase of the active particles, active matter exhibits capillary action even with purely repulsive particle-wall interaction. Contrary to the notion of classical passive fluids, where phenomena such as wall wetting and capillary action originate in wall-liquid adhesive forces and inter-molecular cohesive forces inside the liquid, an active scalar fluid is able to mimic such a behavior in absence of any attractive forces within the system. However, due to the inherent out-of-equilibrium nature of active matter~\cite{fodor2016}, a quantity analogous to surface tension cannot be defined in passive equilibrium systems and hence, the simple intuition underlying capillary action based on the balance between gain in surface energy and gravitational energy of the liquid column fails here. A recent study has shown that self-propelled Janus colloids exhibit unexpected adhesion and alignment of particles at the wall~\cite{wysocki2023}, which enhance the capillary action by enabling active particles to climb up a wall against gravity.

Several studies have shown the presence of stationary particle currents in the context of scalar active matter. Although no alignment mechanism is present, persistent cooperative motion of particles has been observed in the dense phase of ABPs~\cite{wysocki2014}, where an effective velocity alignment is observed in presence of MIPS due to the interplay between steric repulsion and activity~\cite{caprini2020, liao2018}. Active particles arrange in vortex-like geometry with a size increasing with the self-propulsion velocity~\cite{caprini2020}, and dense assembly of polydisperse particles move in irregular turbulent flows~\cite{keta2022}. Recently, similar stationary currents have been observed for motile cells in an isolated ellipsoidal compartment~\cite{cammann2021}, for active microrobots in a box~\cite{scholz2017}, or for ABPs at boundary inhomogeneities~\cite{zakine2020, bendor2022}. In the context of ABPs in a box, a universal relation between the non-equilibrium probability flux of the motion and the global geometric properties, via the boundary's curvature, has even been established~\cite{cammann2021}. 

In this paper, we employ a minimal model of interacting ABPs under gravity inside a two-dimensional rectangular box to characterize the wall-wetting mechanism of an active sedimenting fluid. First we intend to confirm that capillary rise is also present in the ABP system as it has been predicted for the active lattice gas model~\cite{wysocki2020} and to scrutinize quantitative similarities and discrepancies. Then, our main goal is to relate the capillary rise or wall wetting with stationary particle currents in the system and to study, how it varies with the particle-particle interaction strength, down to the ideal, non-interacting case.

The paper is organized as follows. We first describe our model in Sec.~\ref{sec_model} and present a detailed analysis of the density profiles in Sec. \ref{sec_rho}. Sec.~\ref{sec_vortices} contains our results on the characterization of the current field and the vortices. Sec.~\ref{sec_F0} presents the evolution of wetting height and vortices when tuning the particle-particle interaction, and Sec.~\ref{sec_nonint} discusses about our results on non-interacting ABPs. Finally, in Sec.~\ref{sec_con} we conclude with a discussion that elucidates our understanding of the system and proposes future directions.


\section{Model}
\label{sec_model}

Active Brownian particles serve as simple yet powerful tools for modeling the behavior of motile matter in different biological environments. Our model is motivated by experiments on self-propelled colloidal particles sedimenting under gravity~\cite{ginot2015, ginot2018} confined to a two-dimensional plane. We consider $N$ circular, self-propelled, Brownian particles in a 2D box of size $(L_x \times L_y)$ with reflecting boundary conditions along $x$ and $y$ directions, subject to a gravitational force along $-{\bf \hat{y}}$. The particles propel themselves forward with a constant propulsion speed $v_s$ and their orientations perform a rotational diffusion with diffusion constant $D_r$ such that all motion is restricted to the $(x,y)$ plane. The particles are considered to be smooth spheres such that there is no hydrodynamic coupling and interchange of angular momentum leading to systematic torques that might aid alignment interactions. Configuration of the system at each instant of time $t$ is given by the positions and self-propulsion directions $\{ {\bf r}_i(t),\theta_i(t) \}$ of all $N$ particles that obey the following equations, 
\begin{gather}
\dot{{\bf r}_i}=v_s{\bf \hat{e}}_i-v_g{\bf \hat{y}}+\frac{{\bf F}_i}{\gamma}, \label{unscaled1}\\
\dot{\theta_i}= \sqrt{2D_r}\eta.\label{unscaled2}
\end{gather}
The motion of each particle $i$ is governed by a self propulsion velocity of constant magnitude $v_s$ directed along $\hat{\mathbf{e}_i}=(\cos \theta_i, \sin \theta_i)$, the sedimentation velocity $v_g$ due to the gravitational force along $-{\bf \hat{y}}$, and a repulsive interaction force $\mathbf{F}_i$ on the $i^{\rm th}$ particle due to its $m$ neighbors with the drag coefficient $\gamma$. $\eta$ is a Gaussian white noise with zero mean and unit variance.

We consider poly-disperse ABPs with radii $R_i$ uniformly distributed in $[0.4,0.6]$, resulting in a mean diameter of $a=1$. The particles interact repulsively with a spring-like force such that the force exerted on particle $i$ is given by ${\bf F}_i= \sum_{j=1}^{N}{\bf F}_{ij}  + {\bf F}_i^{\rm wall}$, with 
\begin{equation}
{\bf F}_{ij} = \begin{cases} k(R_i +R_j - r_{ij}){\bf \hat{r}}_{ij}, &\forall \ r_{ij} < R_i +R_j \\
0, &\text{otherwise}
\end{cases} \label{F0}
\end{equation}
and ${\bf F}_i^{\rm wall}=-\nabla V^{\rm wall}({\bf r}_i)$ the repulsive particle-wall force derived from truncated Lennard-Jones potentials along the four walls, diverging at $x=0,L_x$, $y=0,L_y$ and with range $R_i$. Note that particle $i$ and $j$ only interact when they overlap, which means their distance $r_{ij}=|{\bf r}_i-{\bf r}_j|$ is smaller than the sum of their radii, $R_i+R_j$. Without any loss of generality, we also choose the unit time as $t_0 = a/v_s = 1$. The global packing fraction is given by $\varphi = \rho_0\sum_i \pi R_i^2/N$, where $\rho_0=N/L_xL_y$ is the global number density of ABPs. 


We define the swimming P\'eclet number of the active particles ${\rm Pe}_s=v_s/a D_r$, the ratio of the sedimentation velocity and the swimming velocity $\alpha=v_g/v_s$, the gravitational P\'eclet number ${\rm Pe}_g=\alpha {\rm Pe}_s$ and the particle-particle repulsion strength $F_0=ka/\gamma v_s$. We choose $F_0$ such that the overlap between adjacent ABPs does not exceed $\sim 1 \%$ of the particle diameter. $L_x$ and $L_y$ are chosen to be larger than all persistence length scales of the system and we also set $L_y \gg L_x$ so that the probability of accumulation of the particles on the upper plate is negligible and the particles sediment on the lower plate forming a dense layer at the bottom with a dilute layer of ABPs on top.

To integrate Eqs.~(\ref{unscaled1})-(\ref{unscaled2}) we employ a forward Euler method with a stepsize $dt=0.001$ which implies that it takes $\tau=1/dt$ steps for each particle to move through a distance of one mean particle diameter. We focus here on the stationary state of the stochastic dynamics defined in Eqs.~(\ref{unscaled1})-(\ref{unscaled2}). We start with randomly distributed ABPs within the box and run the simulation until a stationary state is reached ($t_{\rm eq} \simeq 10^3$). Then, we measure steady state quantities averaged over at least $5000$ configurations with a waiting time of $\Delta t = 1$ between two successive realizations, and over at least $100$ initial distributions. A corresponding video file of the time-evolution of $N=5000$ active particles in a $100 \times 400$ box is attached in the Supplemental Material~\cite{SM} as Movie 1, for the parameters: ${\rm Pe}_s=30$, $\alpha=0.2$ and $F_0=100$. The C++ code used to compute the numerical solutions of Eqs.~\eqref{unscaled1}-\eqref{unscaled2} is available in Ref.~\cite{zenodo}.


\section{Particle and Current Density Profiles}
\label{sec_rho}

\begin{figure}[tb]
\begin{center}
\includegraphics[width=\columnwidth]{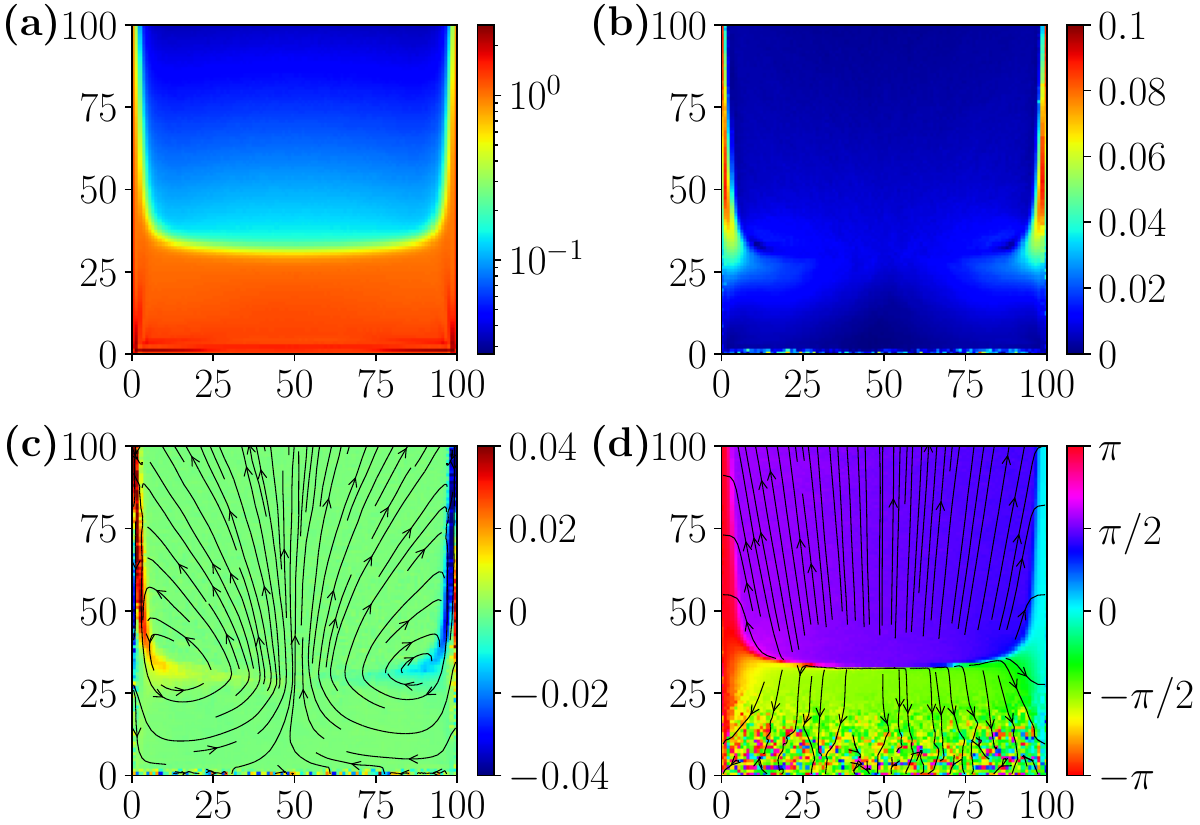}
\caption{Stationary state of ABPs in a box with reflecting walls. Box dimension is $100 \times 400$, particle number is $5000$, gravity is in $-\hat{y}$ direction, $F_0=100$, ${\rm Pe}_s=30$, and ${\rm Pe}_g=6$. Shown quantities are time-averaged. (a)~Particle density $\rho(x,y)$. (b)~Modulus of the current density $|{\bf J}(x,y)|$. (c)~Curl amplitude $A(x,y)$ together with arrows indicating current orientation $\phi_J(x,y)$. (d)~Average particle orientation $\bar \theta(x,y)$.}
\label{denref}
\end{center}
\end{figure}

As can be seen from an exemplary stationary density profile shown in Fig.~\ref{denref}(a), the sedimenting ABPs form a meniscus at the vertical walls, in spite of the repulsive particle-wall interactions. Here, we have considered $N=5000$ active particles in a $100 \times 400$ box -- the global number density is $\rho_0=0.125$ and the global packing fraction is $\varphi \sim 0.098$ -- with the following parameters: $F_0=100$, ${\rm Pe}_s=30$, and ${\rm Pe}_g=6$. This capillary rise, which is absent in passive systems with repulsive particle-wall interactions, emerges due to the propensity of self-propelled particles to accumulate at confining walls in combination with the gravitational force pulling the particles downwards, analogous to what happens in the active lattice gas \cite{wysocki2020}.


A closer look at the particle current in the stationary state reveals the proper mechanism underlying the formation of the meniscus. We define the particle current density ${\bf J}({\bf r})$ as
\begin{gather}
{\bf J}({\bf r}) = \frac{1}{N} \sum_{i=1}^{N} \langle {\bf \dot r}_i(t) \,\delta({\bf r}-{\bf r}_i(t))\rangle_t,
\end{gather}
where ${\bf \dot r}_i(t)$ denotes the velocity of the $i^{\rm th}$ particle and $\langle\cdot\rangle_t$ denotes an average over time and noise. From ${\bf J}({\bf r})$ we extract the time-averaged orientation $\phi_J(x,y)$ via ${\bf J} \propto (\cos \phi_J, \sin \phi_J)$, and the curl amplitude $A(x,y) = \partial_x J_y - \partial_y J_x$. The magnitude and the orientation of the current vector field show a complex structure near the two boundary walls at $x=0$ and $x=L_x$ and the liquid-gas iso-density line, as shown in Figs.~\ref{denref}(b) and~\ref{denref}(c). One sees that the current field due to the ABPs in the wetting layer near the walls is aligned along $-{\bf \hat y}$ direction and away from the walls, the flow field re-aligns in such a way it supports a large vortex near the iso-density line, as indicated by arrows in Fig.~\ref{denref}(c). Note that in ordinary capillary action, particles climb up a wall against gravity, whereas here they appear to climb down instead. This is a consequence of self-propelled particles accumulating at confining walls and the effect of gravitation pulling them down. The flow field is mirror-symmetric about $x=L_x/2$ and one observes two large vortices and curl-clusters concentrated near the left and right boundaries. Thus, contrary to naive expectation, particles do not move upwards along the wall, but downwards close to the wall and upwards -- in a circular current -- at some distance to the wall.

We also calculate the time-averaged polarization vector of ABPs, defined as
\begin{gather}
{\bf P}({\bf r}) = \frac{1}{N} \sum_{i=1}^{N} \langle \hat{\mathbf{e_i}}(t)\,\delta({\bf r}-{\bf r}_i(t))\rangle_t.
\end{gather}
Fig.~\ref{denref}(d) shows the mean-orientation $\bar \theta$ of the ABPs, given by ${\bf P} \propto (\cos \bar \theta, \sin \bar \theta)$. The particles have an effective alignment towards the nearest wall, in the wetting layer and close to the liquid-gas interface (defined further below), from which the wall-accumulation arises. 

In the Supplemental Material~\cite{SM}, we show a view of the entire domain (Fig. S1), a plot of the velocity ${\bf V} = {\bf J}/\rho$ (Fig. S2) and the temporal fluctuations (Fig. S3) of the quantities presented in Fig.~\ref{denref}.

\begin{figure}[tb]
\begin{center}
\includegraphics[width=\columnwidth]{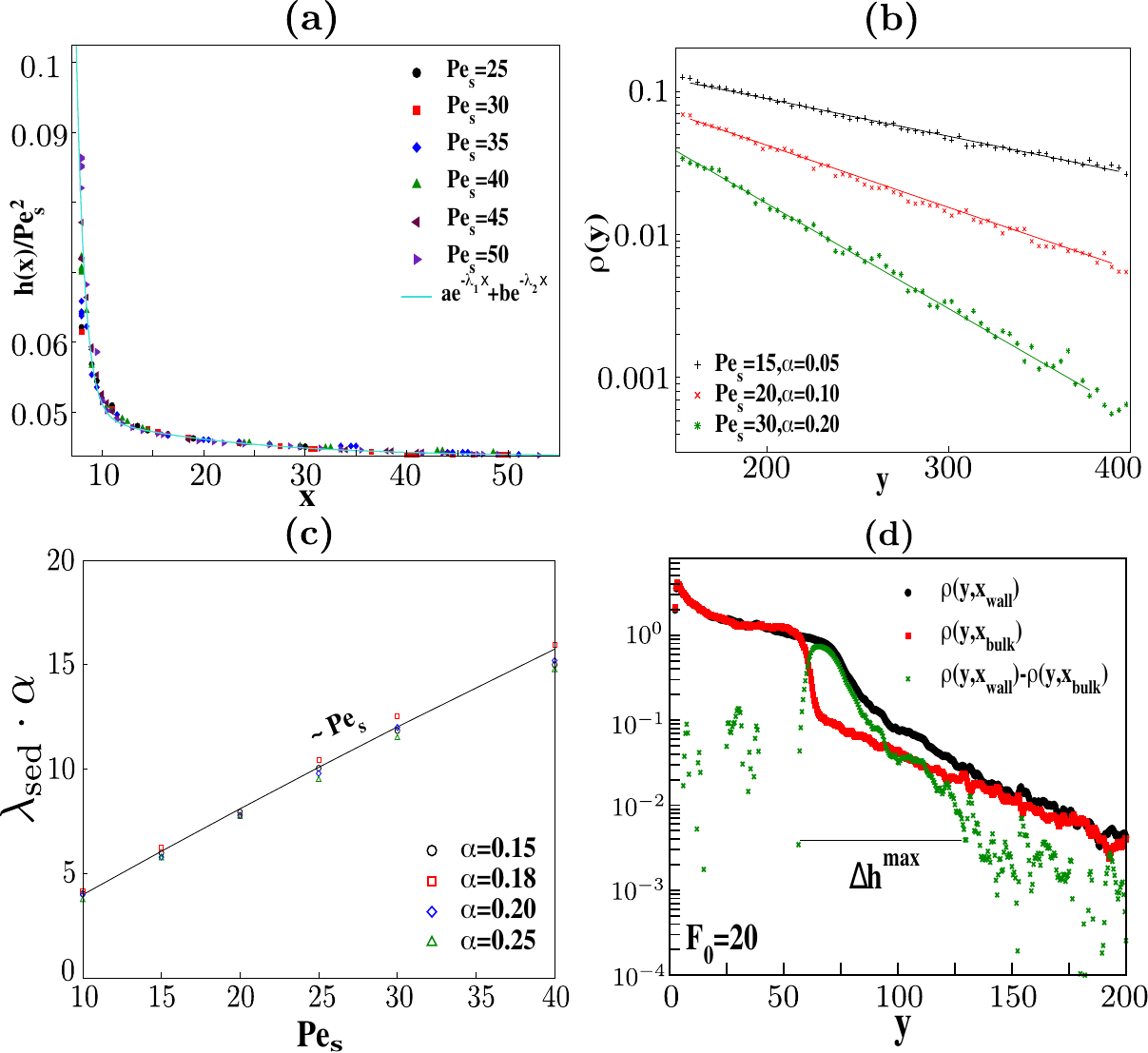}
\caption{(a)~Decay of the iso-density profile $h(x)$, scaled by ${\rm Pe}_s^2$, as a function of distance from the wall $x$, for different ${\rm Pe}_s$ and ${\rm Pe}_g$. The collapsed curves have been fitted to a double-exponential function, shown in solid line. (b)~Density profile $\rho(y) \sim \exp(-y / \lambda_{\rm sed})$ of the dilute layer, for three different sets of (${\rm Pe}_s,\alpha$). (c)~Scaling plot for the sedimentation length extracted from (b): $\lambda_{\rm sed} \sim {\rm Pe}_s^2/{\rm Pe}_g$, such that $\lambda_{\rm sed} \alpha \sim {\rm Pe}_s$ (with $\alpha = {\rm Pe}_s/{\rm Pe}_g$) is linear in ${\rm Pe}_s$. The size of the simulation box is $200 \times 800$ and the particle density is $\rho_0 = 0.125$. (d)~The wall and bulk density profiles $\rho(x_{\rm wall},y)$ (black), $\rho(x_{\rm bulk},y)$ (red) are respectively shown for the parameter set $F_0=20$, ${\rm Pe}_s=50$, and $\alpha=0.4$.  The wetting height $\Delta h^{\rm max}$ is estimated from the difference curve $\rho(y,x_{\rm wall})-\rho(y,x_{\rm bulk})$ (green) from the $y$ values where it decays to a value smaller than $0.01$, as indicated schematically in the plot.} 
\label{fig2}
\end{center}
\end{figure}

To analyze the wetting height, we can consistently define a liquid-gas iso-density interface at $\rho_{\rm iso}= (\rho_l+\rho_g)/2$, where $\rho_l$ and $\rho_g$ are the densities of the dense and dilute phase, respectively. Fig.~\ref{fig2}(a) shows that the scaled height profiles $h(x)$ of the iso-density curves collapse on a master curve, which can be fitted by a double-exponential given by $(h(x)-h_0)/{\rm Pe}_s^2 \sim a \exp(-x/\lambda_1)+b \exp(-x/\lambda_2)$, where $h_0$ is the height of the bulk phase measured with respect to the bottom plate and $x$ is the distance from the wall in the direction transverse to gravity.

The density of the dilute phase decays with vertical distance $y$ from the iso-density line defined above as $\rho(y)\sim \exp(-y/\lambda_{\rm sed})$, as shown in Fig.~\ref{fig2}(b), where $\lambda_{\rm sed}$ is the sedimentation length which scales as $\lambda_{\rm sed} \sim {\rm Pe}_s^2/{\rm Pe}_g$ for large activity ${\rm Pe}_s$~\cite{ginot2015, tailleur2009, solon2015}, as shown in Fig.~\ref{fig2}(c).

In Fig.~\ref{fig2}(d), we plot density profiles $\rho(x_{\rm wall},y)$ and $\rho(x_{\rm bulk},y)$ of the ABPs as a function of $y$, where $x_{\rm wall}=0,L_x$ is situated very close to the left/right walls and $x_{\rm bulk}=L_x/2$ is situated at the middle of the box. At $x=x_{\rm wall}$, the wetting density profile is observed, while at $x=x_{\rm bulk}$, the density profile behaves like in a passive sedimenting system of purely repulsive Brownian particles. We subtract these two densities $\rho(x_{\rm wall},y)$ and $\rho(x_{\rm bulk},y)$ -- the bulk density is expected to decay faster than the wall density -- and the maximum wetting height $\Delta h^{\rm max}$ is measured by estimating the difference between two $y$ values which correspond to $\rho(x_{\rm wall},y)-\rho(x_{\rm bulk},y) \lesssim 0.01$ as shown in Fig.~\ref{fig2}(d).

\begin{figure}[tb]
\begin{center}
\includegraphics[width=\columnwidth]{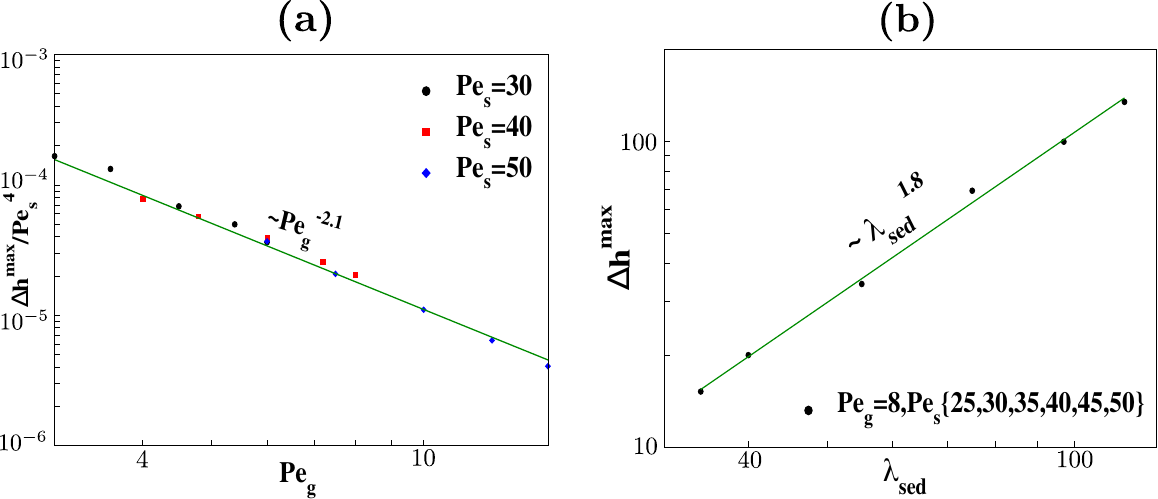}
\caption{(a)~Maximal wetting height $\Delta h^{\rm max}$ as a function of ${\rm Pe}_g$ for different ${\rm Pe}_s$. The scaling $\Delta h^{\rm max} \sim {\rm Pe}_s^4/{\rm Pe}_g^{2.1}$ is shown in solid line. (b)~Maximal wetting height $\Delta h^{\rm max}$ as a function of $\lambda_{\rm sed}$ for different ${\rm Pe}_s$. The scaling $\Delta h^{\rm max} \sim \lambda_{\rm sed}^{1.8}$ is shown in solid line. $\lambda_{\rm sed}$ for each set of $({\rm Pe}_s,{\rm Pe}_g)$ has been estimated separately. Box dimension is $200 \times 500$ and particle number is $2\times 10^4$.}
\label{fig3}
\end{center}
\end{figure}

We further study the dependence of the maximum wetting height $\Delta h^{\rm max}$ of the wetting profiles as a function of ${\rm Pe}_s,{\rm Pe}_g$ and find a scaling behavior $\Delta h^{\rm max} \sim {\rm Pe}_s^{\nu}/{\rm Pe}_g^{\beta}$ with $\nu \sim 4$ and $\beta \sim 2.1$. In Fig.~\ref{fig3}(b) we show $\Delta h_{\rm max}$ as a function of $\lambda_{\rm sed}$ and find a super-linear scaling dependence $\Delta h_{\rm max} \sim \lambda_{\rm sed}^\mu$ with $\mu = 1.8$. Note that this agrees roughly with the scaling reported in Fig.~\ref{fig3}(a) after inserting $\lambda_{\rm sed} \sim {\rm Pe}_s^2/{\rm Pe}_g$. Note that, in a previous study of capillary rise in an ALG setting~\cite{wysocki2020}, the value of the exponent $\mu$ was found to be $1.3$. The wetting properties also depend on the strength of the particle-particle repulsion $F_0$. It turns out that the maximum wetting height decreases with $F_0$ and the meniscus width increases with $F_0$. We discuss our results for varying $F_0$ in Sec~\ref{sec_F0}.


\section{Particle current and vortices}
\label{sec_vortices}


The particle current density ${\bf J}({\bf r})$, depicted in Fig.~\ref{denref}(c), indicates the formation of the meniscus, including its height and width, which is mainly caused by the large circular current -- or vortex -- emerging at the base of the meniscus. Therefore we quantify, in this section, the size and strength of the emerging vortex and its dependence on activity and gravity.

\begin{figure}[tb]
\begin{center}
\includegraphics[width=\columnwidth]{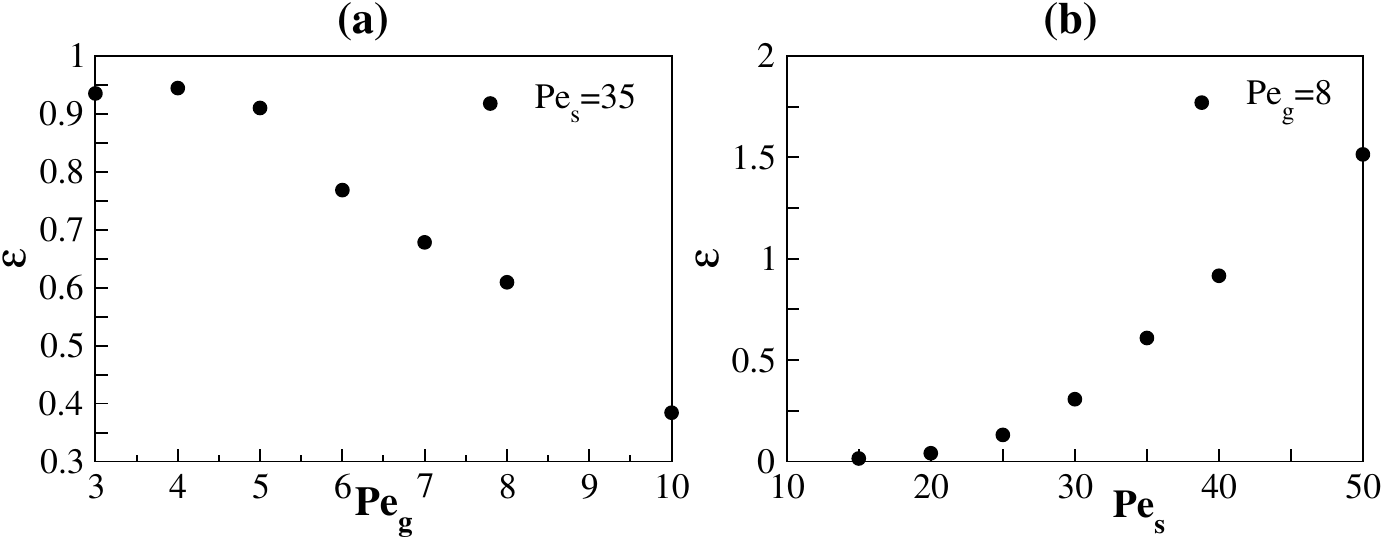}
\end{center}
\caption{(a)~Enstrophy plotted against ${\rm Pe}_g$ for a fixed ${\rm Pe}_s=35$ shows a decrease with increasing gravitational force. (b)~Enstrophy increases as a function of ${\rm Pe}_s$ for fixed ${\rm Pe}_g=8$.}
\label{vortex_stats_Pe}
\end{figure}

First, we calculate the total vorticity in the systems, measured by
the enstrophy
\begin{equation}
\varepsilon = \int_{\rho>\rho_{\rm iso}} dx dy \ | A(x,y) |^2,
\end{equation}
where $A$ is the numerically computed curl of ${\bf J}(x,y)$ and the integral is computed over the entire liquid bulk phase of ABPs below the liquid-gas iso-density line. Fig.~\ref{vortex_stats_Pe} shows that the enstrophy increases with ${\rm Pe}_s$, and decreases with ${\rm Pe}_g$. A major contribution to the vorticity or total curl in the system comes from the shear band along the wall (c.f. Fig.~\ref{denref}(c)), where particles move downwards under the influence of the gravitational force. To quantify the size and strength of the big vortex at the base of the meniscus, alternative methods must then be applied.

\subsection*{Flux line analysis}

\begin{figure}[tb]
\begin{center}
\includegraphics[width=\columnwidth]{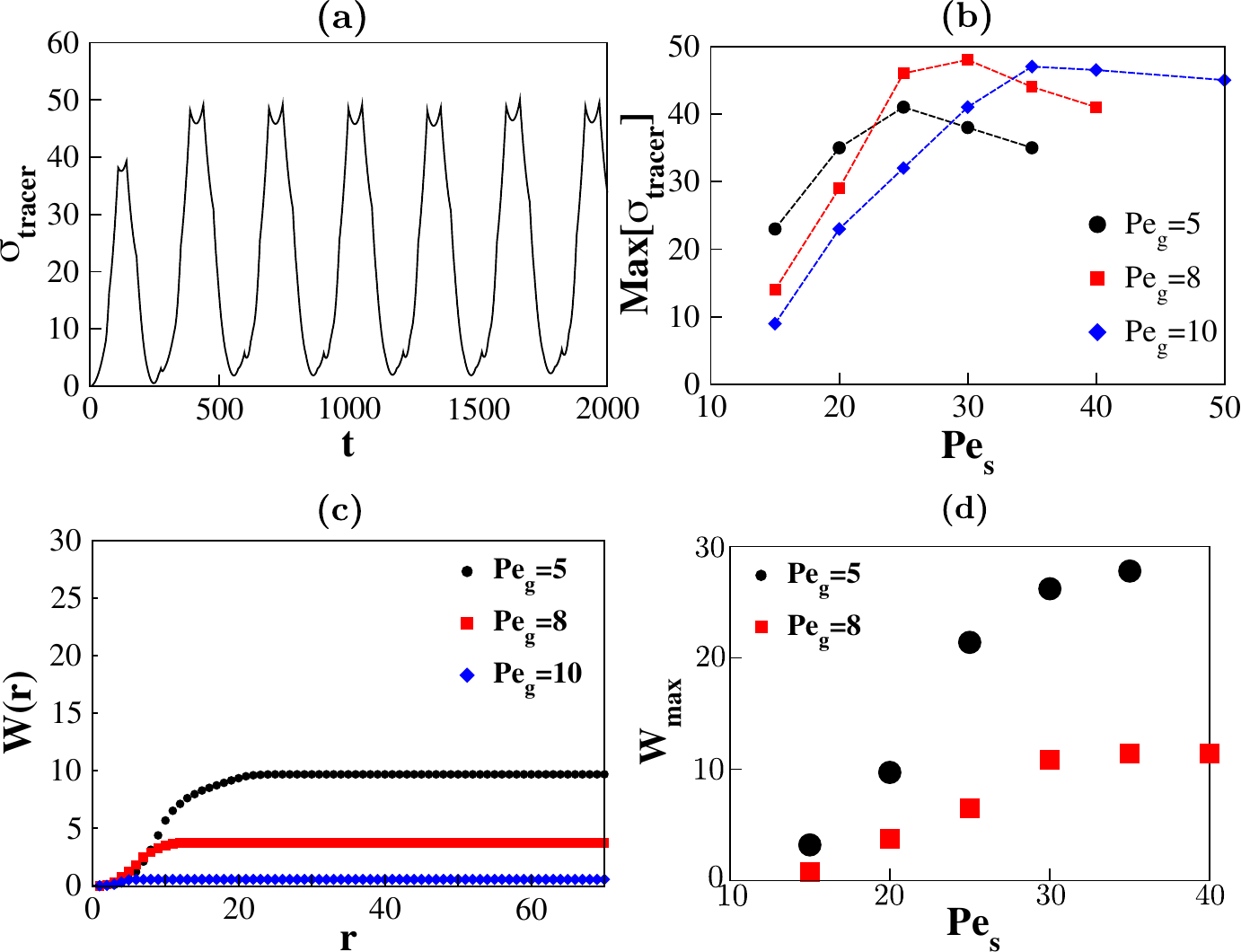}
\end{center}
\caption{(a)~Representative plot for the mean displacement $\sigma_{\rm tracer}$ of a tracer particle on a vortical loop. From the peak value of $\sigma_{\rm tracer}$ one can read off the size of the loop. The largest peak value among all the oscillatory tracer particle trajectories gives the size of the largest vortex. (b)~The largest amplitude in $\sigma_{\rm tracer}$ is plotted as a function of ${\rm Pe}_s$ for three different ${\rm Pe}_g$ values. (c)~Area integral $W(r)=\int_S A dS$, where $S$ is the area of a circle of radius $r$ drawn around the center of the largest vortex. $W(r)$ vs $r$ is plotted for different values of ${\rm Pe}_s,{\rm Pe}_g$. (d) The saturation value of $W(r)$ is plotted against ${\rm Pe}_s$ for two different values of ${\rm Pe}_g$.}
\label{vortex_trajectory}
\end{figure}

First, we consider the trajectories of a virtual tracer particle ${\bf r}_{\rm tracer}(t)$ under the influence of a vector field defined by the current density
\begin{equation}
\label{eqTracer}
\dot{\bf r}_{\rm tracer}={\bf J}({\bf r}_{\rm tracer}),
\end{equation}
With the stationary current density field ${\bf J}$ that we determined above, we integrate numerically the differential equation~\eqref{eqTracer} from a given initial position ${\bf r}_0$. If the initial position ${\bf r}_0$ of a tracer particle lies on a vortex loop, then the mean displacement $\sigma_{\rm tracer}=|{\bf r}_{\rm tracer}(t)-{\bf r}_0|$ shows oscillations as a function of time and one can estimate the size of the loop from the maximum amplitude of $\sigma_{\rm tracer}$, as shown in Fig.~\ref{vortex_trajectory}(a). Note that the period of the oscillations in Fig.~\ref{vortex_trajectory}(a) can be identified with a turn-over time of the vortex, and is around 300 time units for the P\'eclet number considered there, which is slow compared to the velocity of the particles.

We search for the largest closed loop in the velocity field using the maximum amplitude of $\sigma_{\rm tracer}$ as a measure of the mean radius of the two large vortices in the system. We denote the maximal amplitude of the mean displacement of a tracer in the current ${\bf J}({\bf r})$ as $\max[\sigma_{\rm tracer}]$, to provide a first estimate of the spatial extension of the vortex. Fig.~\ref{vortex_trajectory}(b) shows this maximum $\max[\sigma_{\rm tracer}]$ as a function of ${\rm Pe}_s$ and ${\rm Pe}_g$. For a fixed ${\rm Pe}_g$ the mean radius depends non-monotonically on the swimming P\'eclet number ${\rm Pe}_s$, which we can rationalize as follows: the vortex emerges due to the self-propulsion of the particles, for which reason one expects the size and strength of the circular current to increase with swimming P\'eclet number ${\rm Pe}_s$, which is indeed the case for small ${\rm Pe}_s$. However, for larger ${\rm Pe}_s$ the escape probability of the ABPs supersedes the gravitational force such that the outer flow lines of the vortex do not close and hence $\max[\sigma_{\rm tracer}]$ decreases. For even larger ${\rm Pe}_s$ values, one does not find a closed vortex in the flow field. For small ${\rm Pe}_s$, with increasing gravitational force ${\rm Pe}_g$,  the wetting height decreases and the vortex gets more concentrated towards the walls. Consequently, the vortex size decreases with ${\rm Pe}_g$ for lower ${\rm Pe}_s$. However, as ${\rm Pe}_s$ increases, a larger gravitational pull is required for the flow fields to close and give rise to a vortex and hence, for larger ${\rm Pe}_s$, $\max[\sigma_{\rm tracer}]$ increases with ${\rm Pe}_g$. 

Finally, the strength of the vortex can be quantified by an integral over the curl, $W(r)=\int_S A dS$, where $S$ is a circle of radius $r$ around the center of the largest vortex. As shown in Fig.~\ref{vortex_trajectory}(c), $W(r)$ increases with $r$ and saturates at the boundary of the vortex at a value $W_{\rm max}$, which we identify with its strength. Fig.~\ref{vortex_trajectory}(d) shows that the vortex strength increases and saturates with ${\rm Pe}_s$ and decreases with ${\rm Pe}_g$.

\subsection*{Curl cluster analysis}

As a measure of the spatial extent of the vortices, we perform a cluster analysis of the curl amplitude $A(x,y)$. By introducing a small threshold value $A_0=0.01$ for the curl strength, we can identify connected clusters in which all sites have a curl strength larger than this threshold $A({\bf r})>A_0$. The largest curl clusters can be distinguished into two regions: the shear zone close to the wall where the wetting takes place and the other close to the liquid-gas iso-density line close to the left and right boundary walls where the vortices form. The layer of ABPs wetting the wall experiences a repulsion due to the reflecting wall and hence undergoes a slow re-orientation as a result of the collisions with the wall. This mechanism gives rise to a large magnitude of curl in the wetting layer close to the wall. The area of the large cluster in the wetting layer close to the wall is denoted as $S_{\rm wall}$, and the area of the cluster close to the iso-density line (but outside the wetting layer) is denoted $S_{\rm bulk}$, both of which provide an alternative estimate of the spatial extent of the vortices.

\begin{figure}[tb]
\begin{center}
\includegraphics[width=\columnwidth]{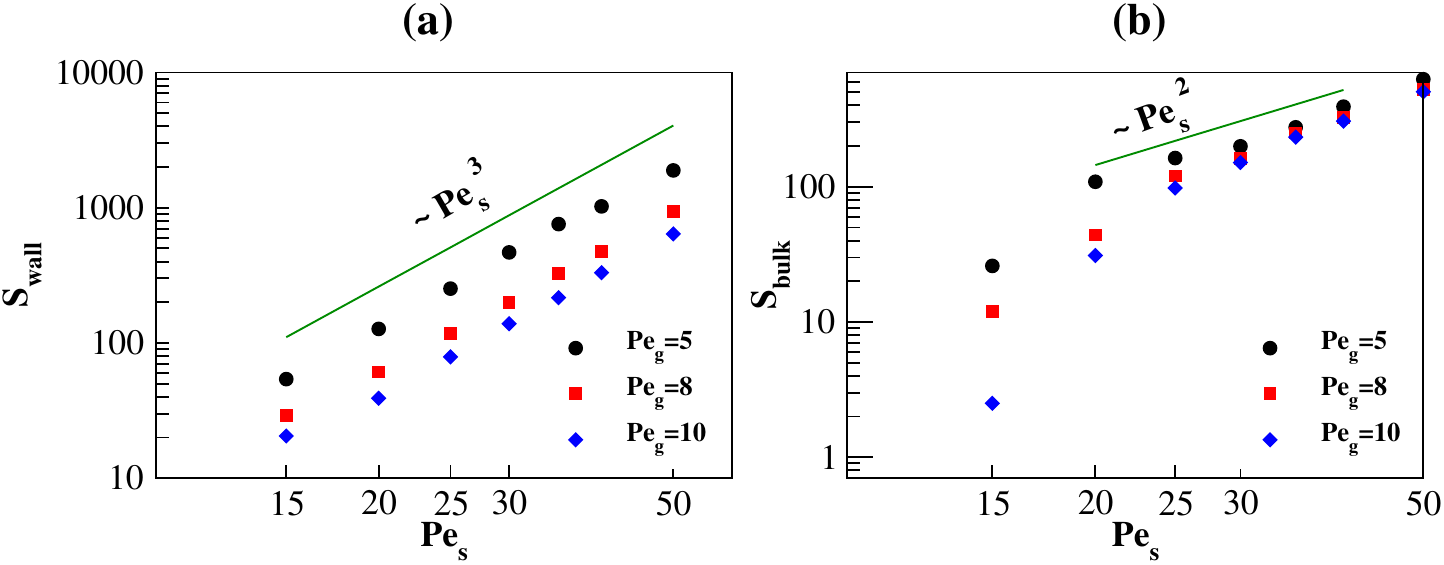}
\end{center}
\caption{Areas of the largest curl clusters (a)~$S_{\rm wall}$ and (b)~$S_{\rm bulk}$ plotted against ${\rm Pe}_s$ for 3 different values of ${\rm Pe}_g$. $S_{\rm wall}$ and $S_{\rm bulk}$ show power-law behavior $ \sim {\rm Pe}_s^\nu$ with $\nu=3$ and $\nu=2$, respectively.}
\label{cluster}
\end{figure}

Fig.~\ref{cluster} shows $S_{\rm wall}$ and $S_{\rm bulk}$ as functions of ${\rm Pe}_s$ for three different values of ${\rm Pe}_g$. The cluster size increases with ${\rm Pe}_s$ since with higher swimming persistence the ABPs wet the walls more (see Fig.~\ref{fig3}) and have a higher escape rate probability from the liquid-gas interface. $S_{\rm wall}$ and $S_{\rm bulk}$ show power-law behavior $ \sim {\rm Pe}_s^\nu$ with $\nu=3$ and $\nu=2$, respectively. However, the cluster size decreases with ${\rm Pe}_g$, due to increased gravitational persistence, the maximum wetting height decreases and so does the escape rate from the liquid-gas iso-density interface. As a consequence, with increasing ${\rm Pe}_g$, the flow gets more concentrated towards the walls, thus decreasing the effective area of the vortices.

\begin{figure}[tb]
\begin{center}
\includegraphics[width=\columnwidth]{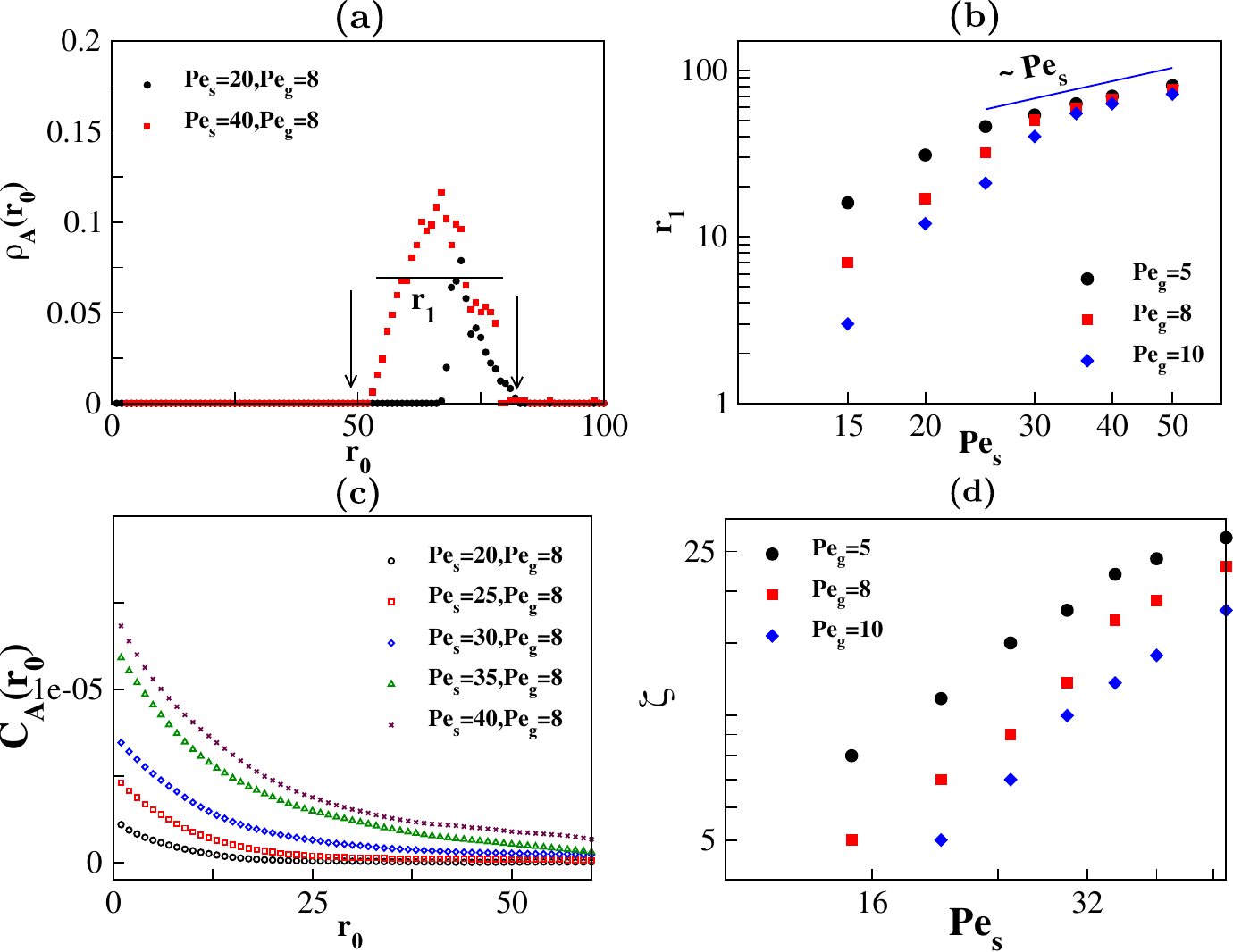}
\end{center}
\caption{(a)~Density $\rho_A$ for the curl variable as a function of the distance $r_0$ measured from the lower corners of the box. Since the density profiles are symmetric about $x=L/2$, we average over the left and right halves of the box for measuring $\rho_A$. (b)~Length scale $r_1$ plotted against ${\rm Pe}_s$ for three different ${\rm Pe}_g$ values. (c)~Two-point correlation function of the curl amplitude $C_A$ plotted against $r_0$ for five sets of ${\rm Pe}_s,{\rm Pe}_g$. (d)~Correlation length $\zeta$ estimated for $C_A$ using an exponential fit, as a function of ${\rm Pe}_s$, for three different values of ${\rm Pe}_g$.}
\label{1pt_corr}
\end{figure}

Furthermore, we measure the density function for the curl amplitude and define a two-state variable $\eta$ such that $\eta(x,y)=1$ for $\vert A(x,y) \vert > 10^{-2}$ and $\eta(x,y)=0$ otherwise. We exclude the wall shear zone from our density calculation so that the high curl values due to the wetting layer do not dominate the signal from the vortices.
Fig.~\ref{1pt_corr}(a) shows the curl density $\rho_A(r_0)=\langle \eta (x,y) \rangle$ as a function of $r_0= \sqrt{x^2+y^2}$. The interval where the curl density remains zero indicates that the curl is very small, implying there is no significant current in this region. The curl density becomes positive at $r_0 \sim 50$, roughly indicating the distance from the bottom corners at which the vortex centers are located. Since the numerical calculation has been carried out by excluding the shear wetting zone, one can interpret the length scale $r_1$ where the curl density is high as an estimate of the linear dimension of the largest curl cluster exclusively due to the vortex. Fig.~\ref{1pt_corr}(b) shows that $r_1$ increases linearly with ${\rm Pe}_s$, in accordance with the quadratic dependence of the vortex area $S_{\rm bulk}$ on ${\rm Pe}_s$ shown in Fig.~\ref{cluster}(b). However, note that there is no non-monotonicity with increasing ${\rm Pe}_s$ similar to that found in the vortex sizes obtained from the tracer particle analysis, as presented in Fig.~\ref{vortex_trajectory}(b). 
Indeed, when the vortex loops do not close, there can be significant curl due to the turbulence in the current field and one ends up observing curl clusters larger than the size of the closed vortices for same set of parameters. 

Finally, we measure the two-point correlation function for the curl $A(x,y)$:
\begin{equation}
C_A(\delta x,\delta y)= \langle A(x_0,y_0) A(x_0+\delta x, y_0+\delta y) \rangle_{x_0,y_0},
\end{equation}
averaged over all space points $(x_0,y_0)$ in the domain. We again exclude the wetting layer such that the length scales of the vortices can be extracted from the density and correlation functions of the flow fields. Fig.~\ref{1pt_corr}(c) shows the two-point correlation function $C_A(r_0)$ as a function of $r_0=\sqrt{\delta x^2+ \delta y^2}$.  We observe short-range correlations which arise only due to the two large vortices that form near the left and right boundary walls. Fig.~\ref{1pt_corr}(d) shows the correlation length $\zeta$ estimated for the correlation function $C_A$ using an exponential fit. $\zeta$ shows a power law dependence on ${\rm Pe}_s$, with an exponent depending weakly on ${\rm Pe}_g$. For ${\rm Pe}_g=5$ we find that $\zeta \sim {\rm Pe}_s$, while for ${\rm Pe}_g=10$ the correlation length follows $\zeta \sim {\rm Pe}_s^{1.2}$.


\section{Interaction strength dependence}
\label{sec_F0}

\begin{figure}[tb]
\begin{center}
\includegraphics[width=\columnwidth]{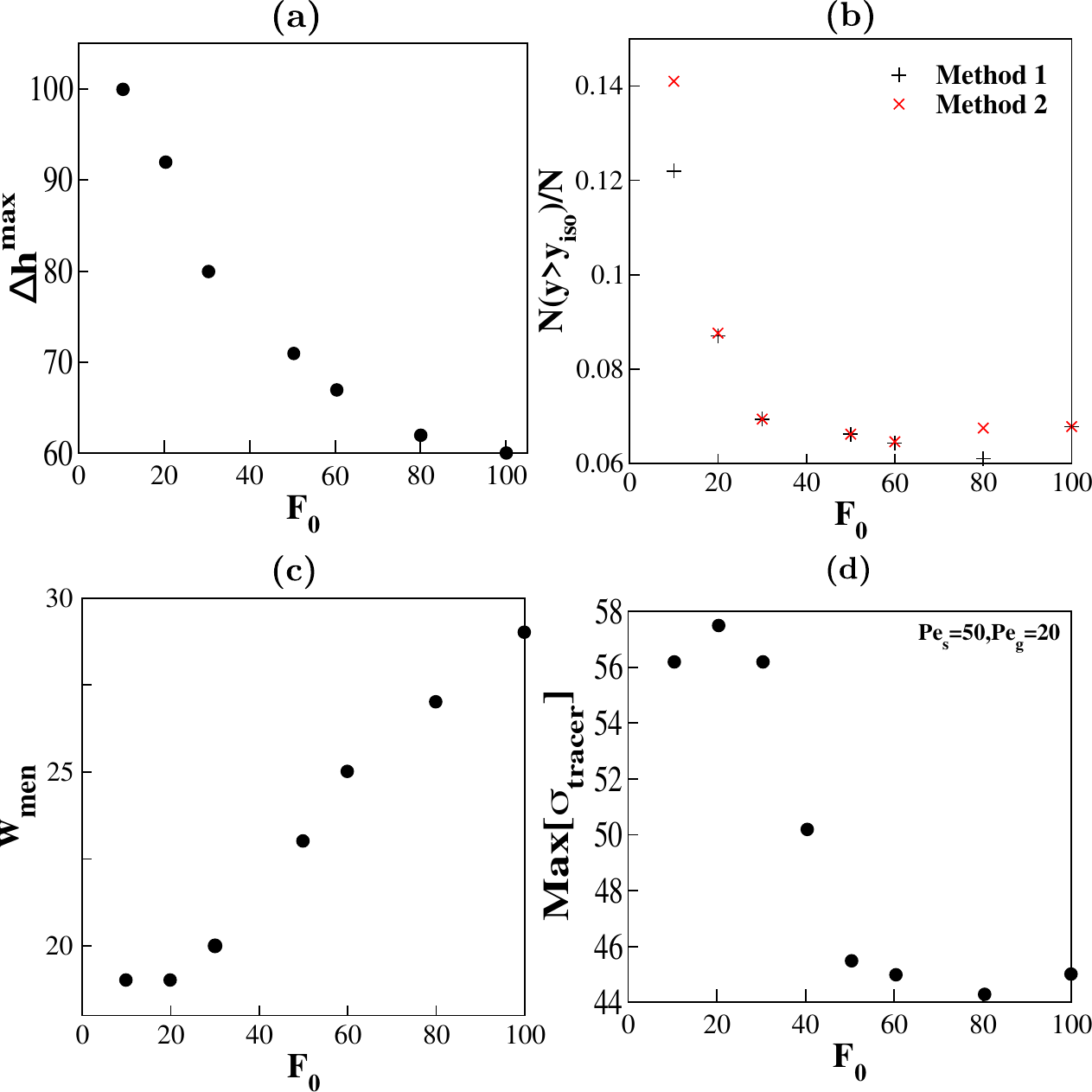}
\caption{(a)~Maximum wetting height $\Delta h^{\rm max}$ plotted against $F_0$ indicates that the wetting height decreases as inter-particle repulsion is increased and approaches a constant value for large $F_0$. (b)~Total number of particles $N(y>y_{\rm iso})$ elevated above iso-density line scaled by the total number of particles in the system plotted against $F_0$. (c)~Width of the meniscus plotted against $F_0$. The width of the meniscus $W_{\rm men}$ increases as a function of $F_0$. (d)~Maximum amplitude of $\sigma_{\rm tracer}$ plotted as a function of $F_0$.}
\label{den_F0}
\end{center}
\end{figure}

Varying the repulsive particle-particle interaction strength $F_0$ changes the effective hard-core diameter of the ABPs and the bulk phase behaves like a more compressible fluid as $F_0$ is decreased. Fig.~\ref{den_F0}(a) shows the maximum wetting height $\Delta h^{\rm max}$ decays with $F_0$ and approaches a constant value for sufficiently large inter-particle repulsion strengths.

We also measure the total number of particles $N(y>y_{\rm iso})$ elevated above the iso-density line scaled by the total number of particles in the system as a function of $F_0$, shown in Fig.~\ref{den_F0}(b). As a consistency check, the height of the bulk iso-density line $y_{\rm iso}$ is estimated using two methods: (1) $\rho_{\rm iso}$ is set to 0.5 uniformly for all $F_0$, (2) Fig.~\ref{fig2}(d) shows two inflection points in $\rho(y)$ that we label as $\rho_l$ and $\rho_g$. For the second method we consider $\rho_{\rm iso}=(\rho_g+\rho_l)/2$. Both methods show that the total mass elevated above $y_{\rm iso}$ decreases with $F_0$. Fig.~\ref{den_F0}(c) shows the width of the meniscus $W_{\rm men}$ plotted against $F_0$ and defined as the smallest distance to the walls where the density $\rho(W_{\rm men},y) \simeq \rho(x_{\rm bulk},y)$. The meniscus width increases as a function of $F_0$. As the ABPs become harder, the wetting layer behaves like an incompressible fluid and hence for nearly equal number fraction of ABPs elevated above $y_{\rm iso}$, the meniscus width increases with $F_0$.

Fig.~\ref{den_F0}(d) shows the size of the vortex also depends on the strength of the particle-particle repulsion $F_0$, for fixed ${\rm Pe}_s$ and ${\rm Pe}_g$. Small values of $F_0$ mean that the particles have a soft-core interaction and the bulk behaves like a compressible fluid, which increases the vortex size in comparison to the less compressible, jammed fluid at large values of $F_0$.

\begin{figure}[tb]
\begin{center}
\includegraphics[width=\columnwidth]{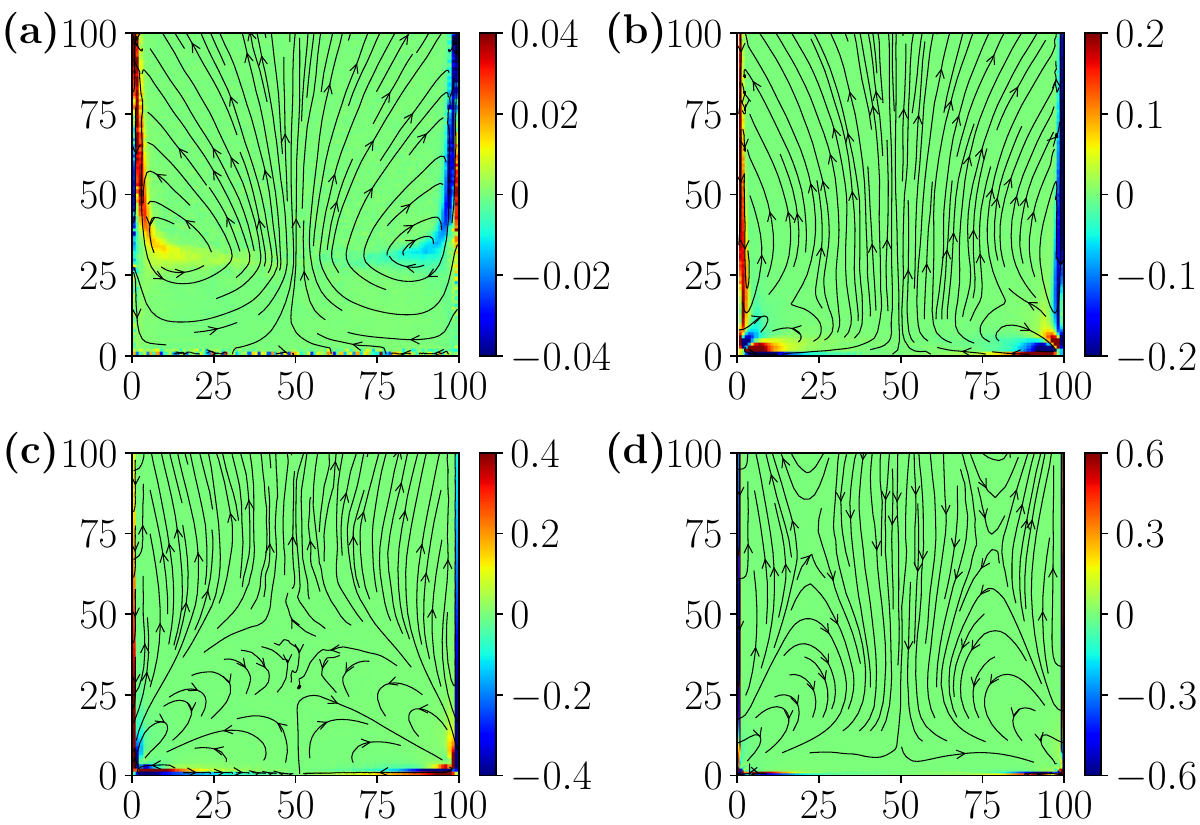}
\caption{Current lines and curl amplitude for fixed ${\rm Pe}_s = 30$ and ${\rm Pe}_g = 6$ ($\alpha=0.2$), and decreasing inter-particle repulsive force: (a)~$F_0=100$, (b)~$F_0=3$, (c)~$F_0=0.3$, and (d)~$F_0=0$. Box dimension is $100 \times 400$, and particle number is $5000$. }
\label{figure_ABP2d_F0}
\end{center}
\end{figure}

Fig.~\ref{figure_ABP2d_F0} shows the evolution of the current density lines and the curl amplitude of the current with decreasing inter-particle repulsive force from $F_0=100$ to $F_0=0$, for fixed ${\rm Pe}_s = 30$ and ${\rm Pe}_g = 6$, in a $100 \times 400$ box. For $F_0 \gtrsim 10$, the inter-particle force is strong enough to create the liquid phase observed in Fig.~\ref{denref}, and the main vortex is located at the base of the meniscus, characterized in Sec.~\ref{sec_vortices}. For $F_0 \lesssim 10$, the particles are no more phase-separated and a system of three vortices is observed at the lower left corner (with two counter-clockwise and one clockwise rotating current) and at the lower right corner (with one counter-clockwise and two clockwise rotating current), characterized in Sec.~\ref{sec_nonint} in the limit $F_0 \to 0$. Hence, the morphology of the stationary particle current and the vortex arrangement changes substantially as soon as MIPS ceases to exist at lower interaction strength $F_0$.


\section{Non-interacting ABPs ($F_0=0$)}
\label{sec_nonint}

The existence of the stationary currents in the system and the vortices in the left and right lower corners can be traced back to the specific geometry of the particle confinement, which can be seen most clearly in the absence of particle-particle interactions, $F_0=0$. For this reason we formulate in this section the hydrodynamic theory for a system of non-interacting active Brownian particles under gravity. Without any interaction between particles, the position and self-propulsion direction of the particles obey the equations
\begin{gather}
\dot {\bf r} = v_s {\bf e_\theta} - v_g {\bf \hat y} + \sqrt{2 D_t} \bm{\eta_r}, \label{ideal1} \\
\dot \theta = \sqrt{2 D_r} \eta_\theta \label{ideal2},
\end{gather}
equivalent to Eqs.~\eqref{unscaled1} and~\eqref{unscaled2} for interacting ABPs. The motion is governed by a self propulsion velocity of constant magnitude $v_s$ directed along ${\bf e_\theta}=(\cos \theta, \sin \theta)$ and the sedimentation velocity $v_g$ due to the gravitational force along $-{\bf \hat{y}}$. $D_t$ and $D_r$ are translational and rotational diffusivities, respectively. $\bm{\eta_r}$ and $\eta_\theta$ are independent Gaussian white noises with zero means and unit variances. From these Langevin equations~\eqref{ideal1} and~\eqref{ideal2}, the probability density function $p({\bf r},\theta;t)$ for a particle to be at position ${\bf r}=(x,y)$ with an orientation $\theta$ at time $t$ follows the Fokker-Planck equation:
\begin{equation}
\label{eq_ideal_cont}
\partial_t p = \nabla \cdot \left[ D_t \nabla p - \left( v_p {\bf e_\theta} - v_g {\bf \hat y} \right) p \right] + D_r \partial_\theta^2 p.
\end{equation}

We numerically solve the steady state of this equation using FreeFEM++~\cite{hecht2013}, a software package based on the finite element method~\cite{zienkiewicz1977}. Writing Eq.~\eqref{eq_ideal_cont} under the form $\partial_t p = - \nabla \cdot {\bf j_r} - \partial_\theta j_\theta$, defining the currents ${\bf j_r} = -D_t \nabla p + ( v_p {\bf e_\theta} - v_g {\bf \hat y}) p$ and $j_\theta = - D_r \partial_\theta p$, the stationary state satisfies the equation
\begin{equation}
\nabla \cdot {\bf j_r} + \partial_\theta j_\theta = 0.
\label{eq_jstat1}
\end{equation}
The weak formulation of Eq.~\eqref{eq_jstat1} is the integral equation:
\begin{equation}
\int_\Omega d{\bf r} d\theta \ w \left( \nabla \cdot {\bf j_r} + \partial_\theta j_\theta \right)  = 0,
\end{equation}
for any arbitrary integrable function $w({\bf r},\theta)$, over a 3d cubic space $\Omega=[-L_x/2,L_x/2] \times [0,L_y] \times [0,2\pi]$. Integrating by part, we have
\begin{equation}
\int_\Omega d{\bf r} d\theta \ \left( \nabla w \cdot {\bf j_r} + \partial_\theta w j_\theta \right)  = 0,
\label{eq_jstat2}
\end{equation}
due to the zero-flux boundary condition in ${\bf x}$ and the periodic boundary condition in $\theta$. This integral equation is solved over the cubic space $\Omega$ divided into a $500 \times 500 \times 16$ tetrahedral mesh-grid. The probability is then calculated at the nodes of the mesh-grid and interpolated linearly over the space with Lagrange polynomials. From this numerical solution for the probability density function $p({\bf r},\theta)$, we extract three integrated functions: the particle density $\rho({\bf r})=\int d\theta \ p({\bf r},\theta)$, the polarization vector ${\bf P}({\bf r}) = \int d\theta \ {\bf e_\theta} p({\bf r},\theta)$, and the current density ${\bf J}({\bf r}) = \int d\theta \ {\bf j_r}({\bf r},\theta)$. With Eq.~\eqref{eq_ideal_cont}, the current density writes
\begin{equation}
{\bf J} = - D_t \nabla \rho + v_p {\bf P} - v_g \rho {\bf \hat y},
\end{equation}
in terms of the density and the polarization vector. We further define the curl amplitude of the current as
\begin{align}
A(x,y) &= \partial_x J_y - \partial_y J_x\nonumber \\
&= v_p \left[ \partial_x m_y - \partial_y m_x \right] - v_g \partial_x \rho.\label{eqCurl}
\end{align}
Without any loss of generality, we set $D_r=1$ and $D_t=1$ defining the scales of time and length, respectively. The remaining parameters are then the swimming P\'eclet number ${\rm Pe}_s = v_s/ \sqrt{D_t D_r}$, the ratio of velocities $\alpha = v_g / v_s = {\rm Pe}_g / {\rm Pe}_s$ and the system size $L_x \times L_y$. The FreeFEM++ code used to compute the numerical solutions is available in Ref.~\cite{zenodo}.

\begin{figure}[t]
\includegraphics[width=\columnwidth]{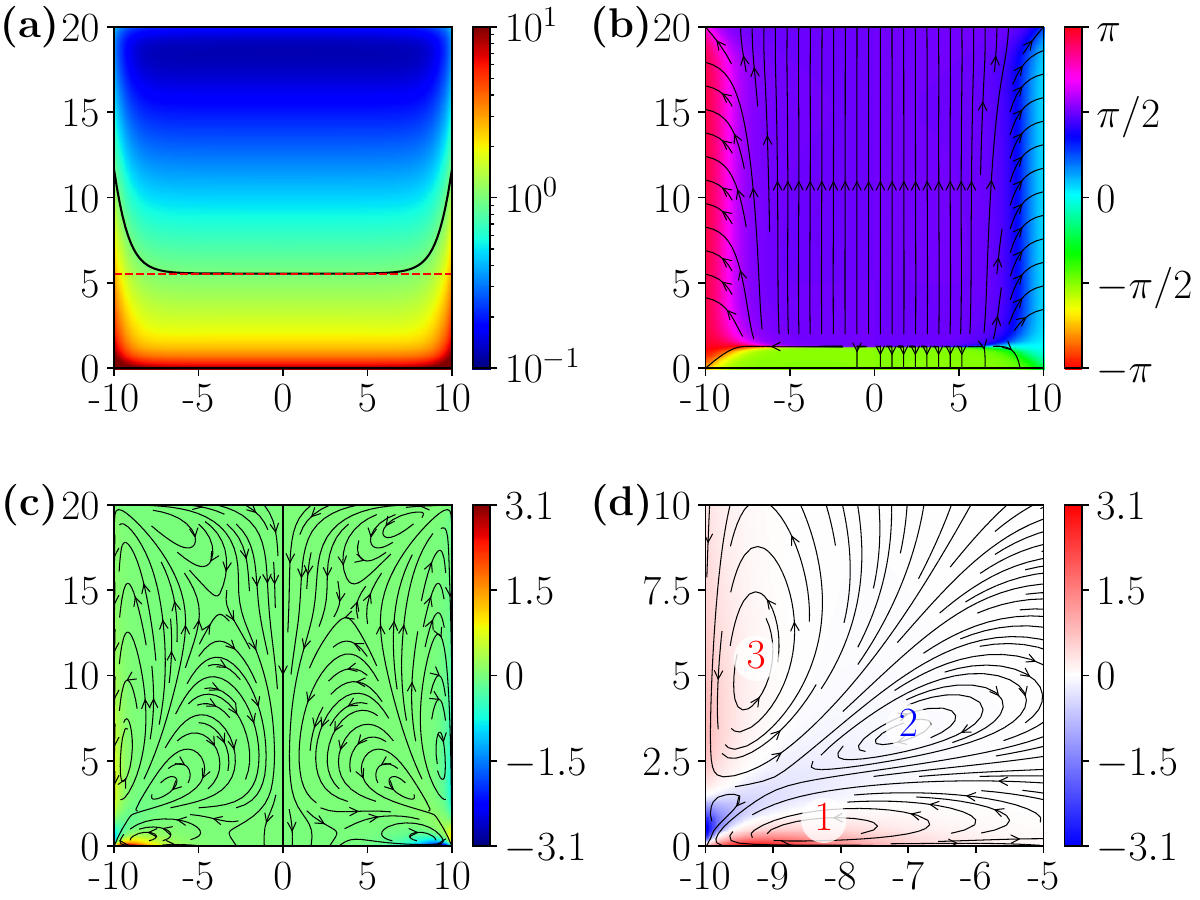}
\caption{Steady state density, polarization and current density profiles for non-interacting ABPs with ${\rm Pe}_s=2$ and $\alpha=0.25$ in a $20 \times 20$ box, obtained numerically with FreeFem++. (a)~Steady state density $\rho(x,y)$. The iso-density line $\rho = 1$ is shown by a solid black line. (b)~Mean orientation $\bar \theta(x,y)$ obtained from the steady state polarization ${\bf P}(x,y)$. (c)~The current density lines $\phi_J(x,y)$ are shown by arrows, and the curl amplitude $A(x,y)$ is represented by the colorbar. (d)~Zoomed-in version of the bottom-left corner of~(c). Three main vortices are observed: vortex 1 and vortex 3 along the bottom and left walls, respectively, and anti-vortex 2 at the corner.\label{figIdealFEM}}
\end{figure}

Fig.~\ref{figIdealFEM} shows numerically obtained steady state density, polarization and current density profiles for non-interacting ABPs with ${\rm Pe}_s=2$ and $\alpha=0.25$ in a $20 \times 20$ box. The density profile shown in Fig.~\ref{figIdealFEM}(a) establishes the existence of a capillary rise near the vertical walls where the particles are mainly oriented towards the wall, as shown in Fig.~\ref{figIdealFEM}(b), with the mean orientation $\bar \theta(x,y)$ calculated from the polarization vector as ${\bf P} \propto (\cos \bar \theta, \sin \bar \theta)$. The wetting height is calculated from iso-density lines, as presented in Fig.~\ref{figIdealFEM}(a) for $\rho=1$ by a solid black line. Despite the absence of any particle interactions, the current field is non-zero and forms vortices at the bottom corners of the box, as shown in Figs.~\ref{figIdealFEM}(c) and~\ref{figIdealFEM}(d), where the current density lines $\phi_J(x,y)$ are calculated from the current as ${\bf J} \propto (\cos \phi_J, \sin \phi_J)$ and the curl amplitude $A(x,y)$ is calculated with Eq.~\eqref{eqCurl}. Note that the current field predicted by the Fokker-Planck equation for ideal ABPs agrees with the one obtained for the microscopic model with interaction $F_0=0$, as shown in Fig.~\ref{figure_ABP2d_F0}(d). Fig.~\ref{figIdealFEM}(d) depicts the presence of three main vortices at the bottom-left corner, with two counter-clockwise rotating currents along the bottom and left walls, and one clockwise rotating current at the corner.

\begin{figure}[t]
\includegraphics[width=\columnwidth]{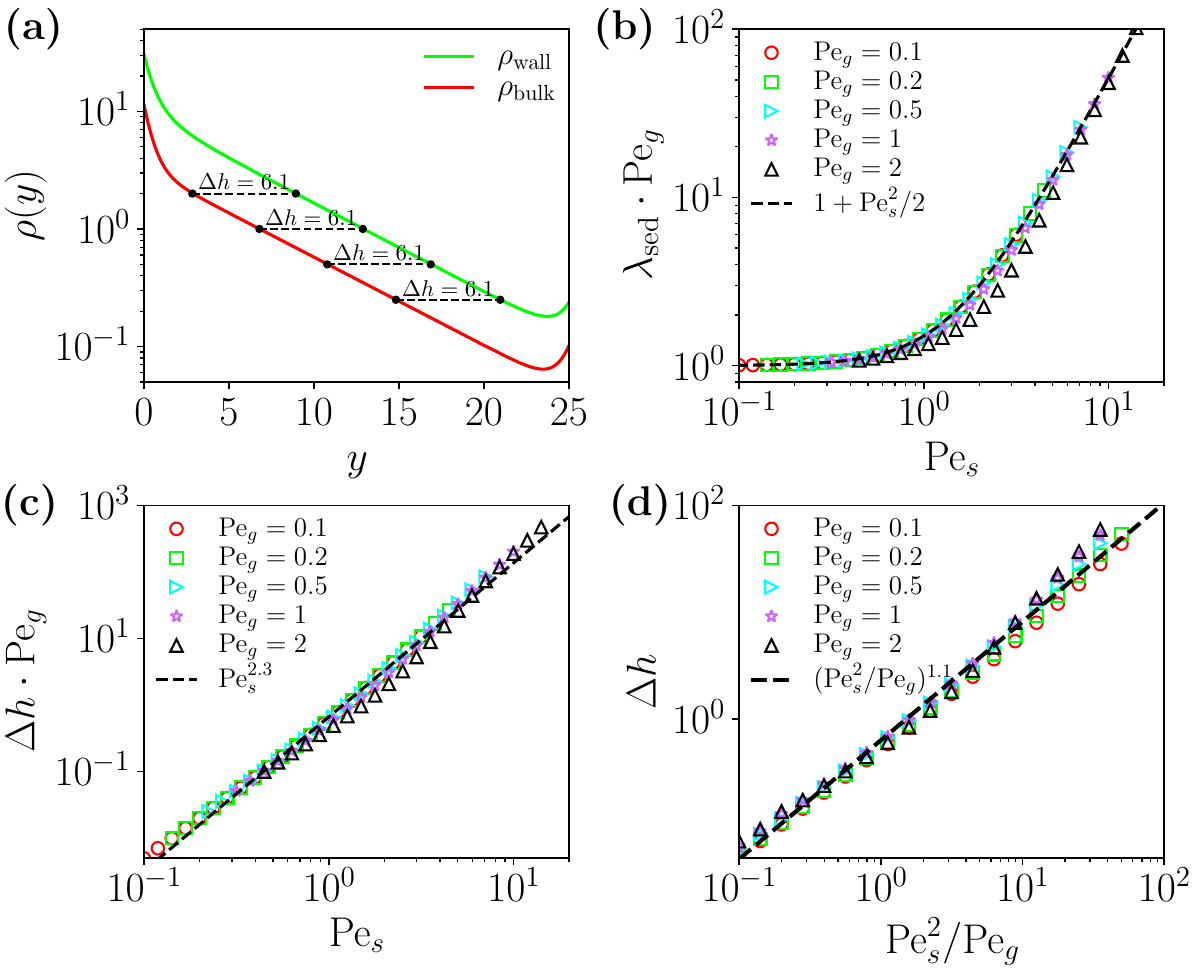}
\caption{(a)~Density profile at the center of the box $\rho_{\rm bulk}$ and near the vertical wall $\rho_{\rm wall}$ for non-interacting ABPs with ${\rm Pe}_s = 2$ and $\alpha=0.25$ in a $25 \times 25$ box, obtained numerically with FreeFem++. For an iso-density line chosen in the exponential decay regime, the wetting height is always $\Delta h = 6.1$. (b)~Sedimentation length $\lambda_{\rm sed}$ as a function of ${\rm Pe}_s$, for several ${\rm Pe}_g$. (c)~and (d)~Wetting height $\Delta h$ as a function of ${\rm Pe}_s$ and ${\rm Pe}_s^2 / {\rm Pe}_g$, respectively, for several ${\rm Pe}_g$. It is calculated with the iso-density line $\rho=1$ in a $100 \times 100$ box. The dotted line represents the fitted curve for all presented data.\label{figIdealRise}}
\end{figure}

Far from the top and bottom walls, the density writes
\begin{equation}
\rho(x,y) = f(x) \exp(-y/\lambda_{\rm sed}) \label{eqExpDecay}
\end{equation}
with the sedimentation length $\lambda_{\rm sed} \simeq {\rm Pe}_s^2 / {\rm Pe}_g$, in the limit of large swimming P\'eclet numbers~\cite{ginot2018}. Fig.~\ref{figIdealRise}(a) shows the density at the center of the box $\rho_{\rm bulk}(y) = \rho(0,y)$ and near the vertical wall $\rho_{\rm wall}(y) = \rho(\pm L_x/2,y)$ with ${\rm Pe}_s = 2$ and $\alpha=0.25$ in a $25 \times 25$ box. They present an exponential decay regime, as expected in the middle region, and the wetting height is then independent of the choice of the iso-density line for a dilute system. Fig.~\ref{figIdealRise}(b) shows the sedimentation length $\lambda_{\rm sed}$ as a function of ${\rm Pe}_s$ for several ${\rm Pe}_g$. In the small and large ${\rm Pe}_s$ limits, we obtain the asymptotic expressions: $\lambda_{\rm sed} \sim 1/{\rm Pe}_g$ and $\lambda_{\rm sed} \sim {\rm Pe}_s^2/2{\rm Pe}_g$, respectively. Merging these two limits, the sedimentation length can be approximated by
\begin{equation}
\lambda_{\rm sed} \simeq \frac{1+0.5{\rm Pe}_s^2}{{\rm Pe}_g},
\end{equation}
shown in Fig.~\ref{figIdealRise}(b) with dashed line, and valid for not too large ${\rm Pe}_g$. Defining $f_{\rm bulk} = f(0)$ and $f_{\rm wall} = f(\pm L_x/2)$ in Eq.~\eqref{eqExpDecay}, the wetting height writes
\begin{equation}
\Delta h = \lambda_{\rm sed} \ln(f_{\rm wall}/f_{\rm bulk}).
\end{equation}
Figs.~\ref{figIdealRise}(c) and~\ref{figIdealRise}(d) show the wetting height $\Delta h$ as a function of ${\rm Pe}_s$ and ${\rm Pe}_s^2 / {\rm Pe}_g$, respectively, for several gravity, and calculated for the iso-density line $\rho=1$ in a $100 \times 100$ box. The wetting height follows the power-laws: $\Delta h \sim {\rm Pe}_s^{2.3} / {\rm Pe}_g$, from Fig.~\ref{figIdealRise}(c), and
\begin{equation}
\Delta h \sim ({\rm Pe}_s^2 / {\rm Pe}_g)^{1.1} \sim \lambda_{\rm sed}^{1.1},
\end{equation}
from Fig.~\ref{figIdealRise}(d), which are both equivalent. This power-law is different from the result obtained for interacting ABPs where $\Delta h \sim \lambda_{\rm sed}^{1.8}$, meaning that the interactions between particles increase the wetting of particles on vertical walls.

\begin{figure}[t]
\includegraphics[width=\columnwidth]{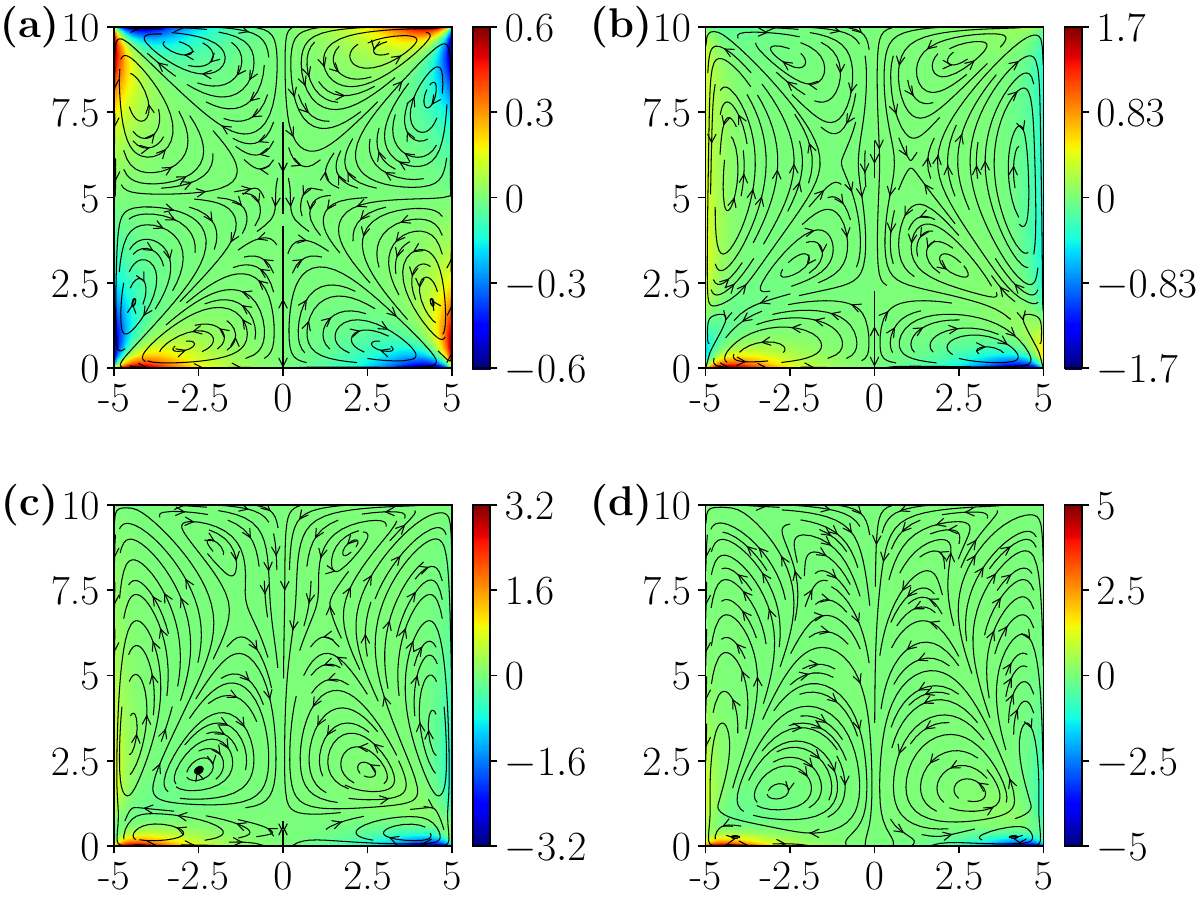}
\caption{Current density lines $\phi_J(x,y)$ for non-interacting ABPs with ${\rm Pe}_s=2$ and (a)~$\alpha=0$, (b)~$\alpha=0.25$, (c)~$\alpha=0.5$, and (d)~$\alpha=0.75$, numerically calculated with FreeFEM++ in a $10 \times 10$ box. The colorbar represents the curl amplitude of the current $A(x,y)$. \label{figIdealVortices}}
\end{figure}

Fig.~\ref{figIdealVortices} shows the current density lines and the curl amplitude of the current with ${\rm Pe}_s=2$ and increasing gravity from $\alpha=0$ to $\alpha=0.75$, in a $10 \times 10$ box. Without gravity, Fig.~\ref{figIdealVortices}(a), the currents self-organize in a way that is compatible with maximum particle accumulation in the corners: incoming flux along the diagonal and outgoing flux parallel to the wall. This eight vortices structure is fully determined by the boundary's geometry, similar to what has been observed for ABPs in an elliptical geometry~\cite{cammann2021}. Under gravity, the two vortices in the upper left and right corner are pulled down, resulting a) in a {\it downward} flux along the vertical walls, and b) in big counter-clockwise and clockwise rotating currents at the bottom close to the lower left and right corners, c.f. Fig.~\ref{figIdealFEM}(d). In addition, the curl amplitude increases with gravity despite the extension of the vortices decrease. Two corresponding video files are attached in the Supplemental Material~\cite{SM} as Movie 2a and Movie 2b, showing the evolution of the particle density and current density lines, respectively, under increasing gravity.

\begin{figure}[t]
\includegraphics[width=\columnwidth]{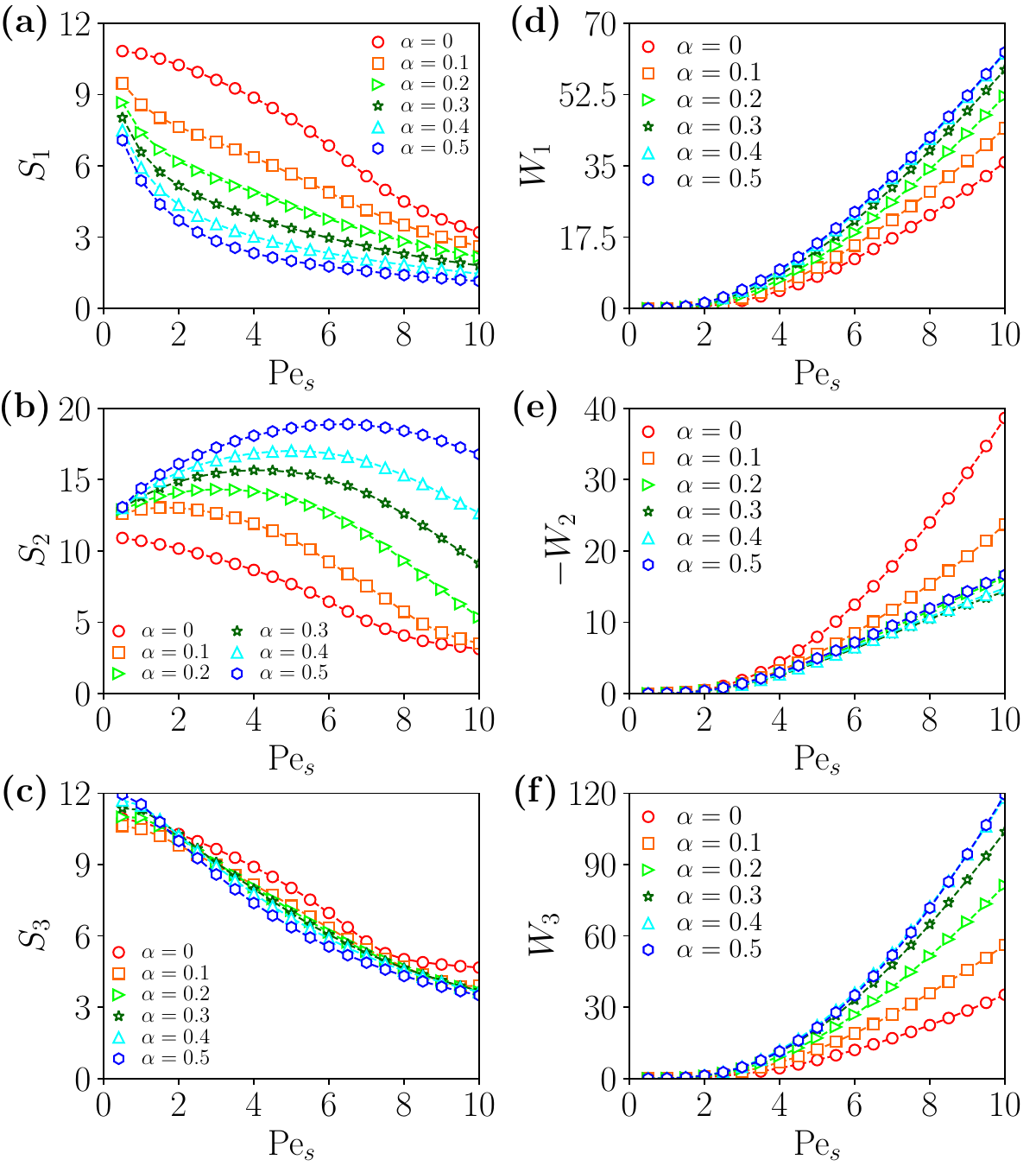}
\caption{(a)-(c)~Vortex area $S_i$ as a function of ${\rm Pe}_s$ for non-interacting ABPs with several $\alpha$ and in a $10 \times 10$ box. (d)-(f)~Circulation $W_i=\int_{S_i} A dS$ of the corresponding vortices. The labels 1, 2 and 3 correspond to the vortices described on Fig.~\ref{figIdealFEM}(d).\label{figIdealArea}}
\end{figure}

Figs.~\ref{figIdealArea}(a)-(c) show the vortex area of the three vortices described on Fig.~\ref{figIdealFEM}(d), as a function of ${\rm Pe}_s$ and $\alpha$, in a $10 \times 10$ box. Fig.~\ref{figIdealArea}(a) shows the vortex area $S_1$ of the counter-clockwise rotating current near the bottom wall. The area is calculated such that the curl amplitude satisfies $|A| > 0.001 A_{\rm max}$, where $A_{\rm max}$ is the local maximum of $|A|$. This vortex area decreases with swimming P\'eclet number and gravity. Fig.~\ref{figIdealArea}(b) shows the vortex area $S_2$ of the clockwise rotating current at the corner. This vortex area has non-monotonous evolution with swimming P\'eclet number, but increases with gravity. Fig.~\ref{figIdealArea}(c) shows the vortex area $S_3$ of the counter-clockwise rotating current near the left wall. This vortex area decreases with swimming P\'eclet number and the gravity has low impact on it. Without gravity, i.e. $\alpha=0$, the current lines of these vortices are anti-symmetric (see Fig.~\ref{figIdealVortices}(a)) and then $S_1 = S_2 = S_3 \lesssim L_x L_y / 8$. This vortex area decreases with swimming P\'eclet number. When the gravity is increased, the current lines are deformed in the $-{\bf \hat y}$ direction, telling that $S_1$ decreases and $S_2$ increases with gravity, while $S_3$ is globally not impacted. $S_1$ and $S_3$ remain decreasing functions of ${\rm Pe}_s$ when $S_2$ has a non-monotonic behavior with ${\rm Pe}_s$.

Figs.~\ref{figIdealArea}(d)-(f) show the circulation of the corresponding vortex, calculated as $W_i=\int_{S_i} A dS$ calculated over the area $S_i$. The circulation of the vortices along the bottom and left walls, $W_1$ and $W_3$ respectively, increases with both ${\rm Pe}_s$ and ${\rm Pe}_g$, as shown in Figs.~\ref{figIdealArea}(d) and~\ref{figIdealArea}(f), respectively. Similarly, the absolute circulation of the anti-vortex $W_2$ increases with ${\rm Pe}_s$, but has a non-monotonic behavior with $\alpha$, as shown in Fig.~\ref{figIdealArea}(e). 

\section{Discussion}
\label{sec_con}

We have shown that a system of active Brownian particles in the phase separated (or MIPS) phase, which sediment in a homogeneous force field, form a wetting meniscus at a confining wall, in spite of repulsive particle-wall interactions. Increasing the activity, measured by the swimming P\'eclet number ${\rm Pe}_s=v_s/aD_r$, increases the height of the meniscus $\Delta h$, and increasing force field, measured by the gravitational P\'eclet number ${\rm Pe_g}= (v_g/v_s) {\rm Pe}_s$, decreases the meniscus height. Quantitatively, $\Delta h$ grows monotonously with the sedimentation length $\lambda_{\rm sed} \sim {\rm Pe}_s^2/{\rm Pe}_g$, approximately like $\Delta h\propto\lambda_{\rm sed}^2$ for strongly repelling particles and roughly linear with $\lambda_{\rm sed}$ for non-interacting ABPs. We also find a non-trivial dependence of the meniscus dimensions on the particle interaction strength or particle softness: softer particles (decreasing $F_0$) increase the meniscus height and decrease the meniscus width, but increase the total elevated mass.

The formation of the meniscus is determined by the formation of a circular particle current, a vortex, centered at the base of the meniscus, which can easily be seen in the movie for interacting ABPs in the Supplemental Material~\cite{SM}, as Movie 1: in the gas region above the iso-density line close to the walls there is a net particle current towards the wall. Particles colliding with the wall stay accumulate there and start to sink towards the liquid region due to the force field. This produces a strong downward particle flow along the wall, which then gets deflected away from the wall when it hits the denser liquid region. Thus, below the liquid-gas interface particles flow away from the wall, and particles reaching the interface have an upward orientation and the circular current closes. Note that there are two walls and therefore two vortices: one in the lower right and one in the lower left corner.

The flow lines of the particle current indicate that each vortex extends over one half of the system, c.f. Fig.~\ref{denref}(c), but it is strongest close to the wall and the meniscus, and very weak far away from it, see Fig.~\ref{denref}(b). It turns out that the total strength of the vortex, measured by its total curl in a concentric disk, increases monotonously with the activity / the swimming P\'eclet number ${\rm Pe}_s$. Analogously the region in which the current is strongest increases with the ${\rm Pe}_s$, too, such that the system's activity determines the size and strength of the circular particle current together with the meniscus or wetting height determined by it.

Interestingly, the origin of the two major vortices can be traced back to the presence of the confining walls of the system: non-interacting (ideal) ABPs in a quadratic area with repulsive walls form stationary probability currents already without a force field. Those are organized according to the fourfold symmetry of the system, namely in each quadrant two vortices, one above and one below the diagonal emanating from the corner. The circulation of each vortex pair in one quadrant is such that the current along the diagonal is directed towards the corner, leading to the well-known accumulation of self-propelled particles in corners or regions, where boundary curvature is high~\cite{cammann2021,ostapenko2018,fily2014}. Switching on the force field (gravity) breaks the fourfold symmetry, squeezes the vortices in the lower quadrants and expands those in the upper half. The two elongated vortices at the walls in the ideal ABP system, one counter-clockwise at the left wall, one clockwise rotating at the right wall, are those that have their counterpart in the interacting ABP system described above.

Concerning the experimental observability of what we have reported in this paper we would like to note that recently the capillary rise along (or “active wetting” of) a wall in a system of active colloids under a gravitational force has been reported~\cite{wysocki2023} and also the emerging particle currents have been discussed. So, in principle the original prediction of~\cite{wysocki2020} as well as what we have reported here have been experimentally confirmed. A few differences should be noted, though: first, the particle activity reached in~\cite{wysocki2023} was, for experimental reasons, much lower than the activities considered here. Therefore, their system was sedimenting but gaseous (i.e. not in the MIPS region). Second, the observed meniscus (or wetting layer) was much thinner than what we obtained here, even thinner than what we report for the ideal (non-interacting, and thus also gaseous) ABP case, and the meniscus height was much larger, i.e. the particles at the wall went much higher above the iso-density line. Third, the experimentally observed particle current along the wall was directed upwards, consistent with the larger meniscus height and giving rise to a clockwise rotating vortex, differing from the downward wall-current and the counter-clockwise rotation reported here. The latter two observations were attributed to additional particle-wall adhesion and alignment forces~\cite{wysocki2023}, both emerging due to hydrodynamic particle-wall interactions of the active colloids. It turned out that the inclusion of those additional particle-wall interactions in an ABP model like the one we considered here could even quantitatively recapitulate the experimental observations.

Finally, the fact that one observes something that is reminiscent of capillary rise at a wall, in spite of repulsive particle wall interactions, is the most obvious signature for the non-equilibrium character of this system. More fundamentally, being out-of-equilibrium in the stationary state implies the presence of stationary probability currents (since otherwise detailed balance would be fulfilled), but these generally live in high-dimensional configuration space. The system we studied here actually shows emergent probability currents leading directly to real-space currents, similar to what has been reported for self-propelled particles in an ellipsoid geometry~\cite{cammann2021}, or for ABPs at boundary inhomogeneities~\cite{zakine2020,bendor2022}. In addition, these currents perform real work by lifting a fraction of the particle mass above the liquid-gas interface against the force field. Thus, one would expect the size and strength of the emerging currents to be related to the entropy production rate of this system, or at least an equivalent one in which the dynamical rules have been defined thermodynamically consistent~\cite{pietzonka2018}. These questions as well as in how far the strength of the observed vortices is related to the vorticity introduced in~\cite{obyrne2023} are interesting and would be worthwhile to be studied in the future.

\begin{acknowledgments}

This work was performed with financial support from the German Research Foundation(DFG) within the Collaborative Research Center SFB 1027-A3 and INST 256/539-1. M.M. wants to thank Dr. Swarnajit Chatterjee for valuable discussions.

\end{acknowledgments}


\clearpage

\onecolumngrid

\setcounter{equation}{0}
\setcounter{figure}{0}
\renewcommand{\theequation}{S\arabic{equation}}
\renewcommand{\thefigure}{S\arabic{figure}}

\begin{center}
{\large \bfseries Supplemental Material for ``Stationary particle currents in sedimenting active matter wetting a wall''}

\bigskip

{\normalsize Matthieu Mangeat,$^1$ Shauri Chakraborty,$^1$ Adam Wysocki,$^1$ and Heiko Rieger$^{1,2}$}

\medskip

{\normalsize \itshape $^1$Center for Biophysics \& Department for Theoretical Physics, Saarland University, D-66123 Saarbr{\"u}cken, Germany.\\
$^2$INM – Leibniz Institute for New Materials, Campus D2 2, D-66123 Saarbr{\"u}cken, Germany.}
\end{center}

\vspace{-0.5cm}

\begin{figure}[H]
\begin{center}
\includegraphics[width=16cm]{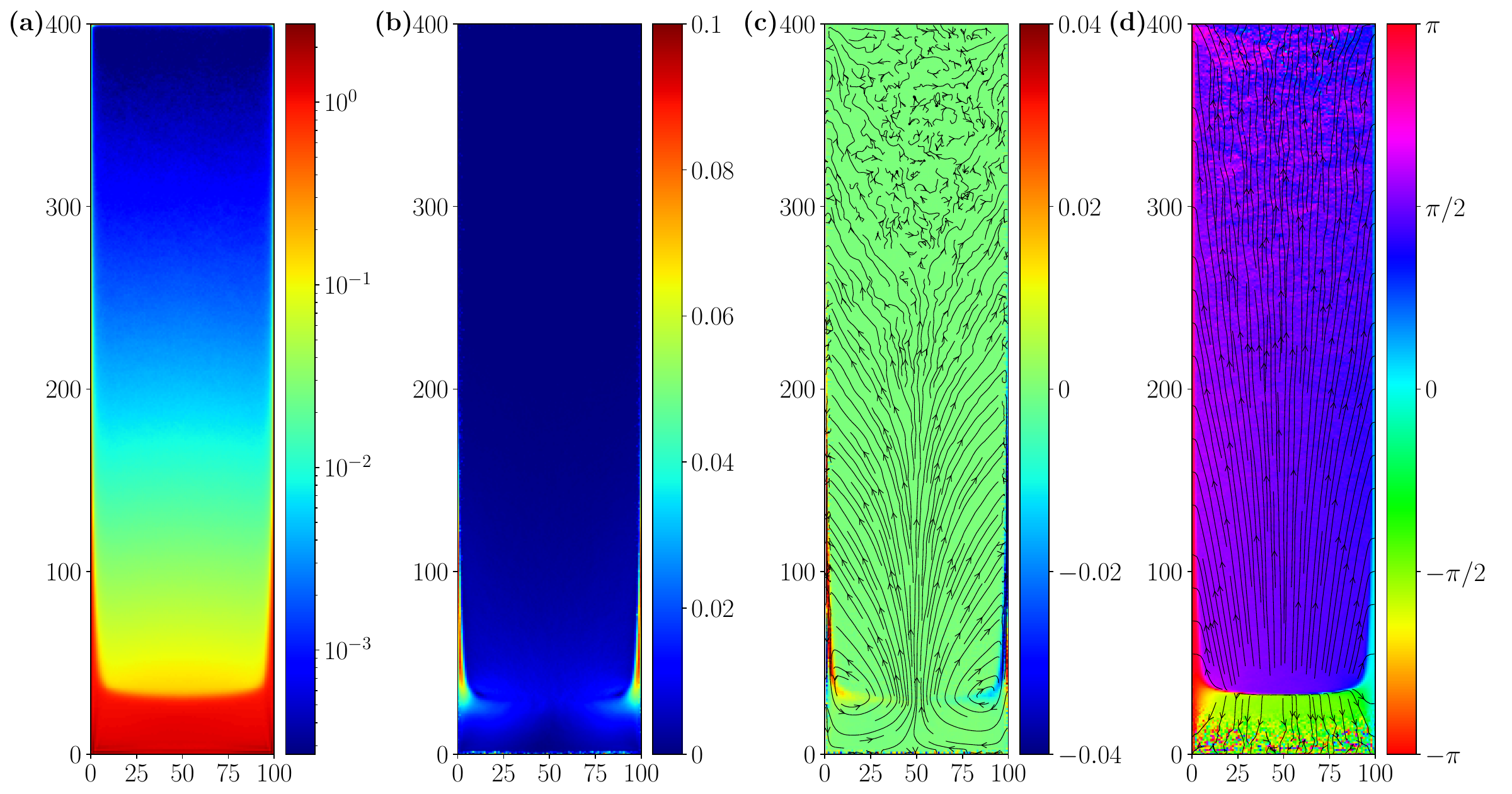}
\caption{[Reproduction of Fig. 1 with a view of the entire domain] Stationary state of ABPs in a box with reflecting walls. Box dimension is $100 \times 400$, particle number is $5000$, gravity is in $-\hat{y}$ direction, $F_0=100$, ${\rm Pe}_s=30$, and ${\rm Pe}_g=6$. Shown quantities are time-averaged. (a)~Particle density $\rho(x,y)$. (b)~Modulus of the current density $|{\bf J}(x,y)|$. (c)~Curl amplitude $A(x,y)$ together with arrows indicating current orientation $\phi_J(x,y)$. (d)~Average particle orientation $\bar \theta(x,y)$. Quantities become more noisy with increasing $y$, due to the lack of particles for a fixed time-averaging window (density presents an exponential decay with $y$ in the dilute region).}
\end{center}
\end{figure}

\vspace{-0.5cm}

\begin{figure}[H]
\begin{center}
\includegraphics[width=16cm]{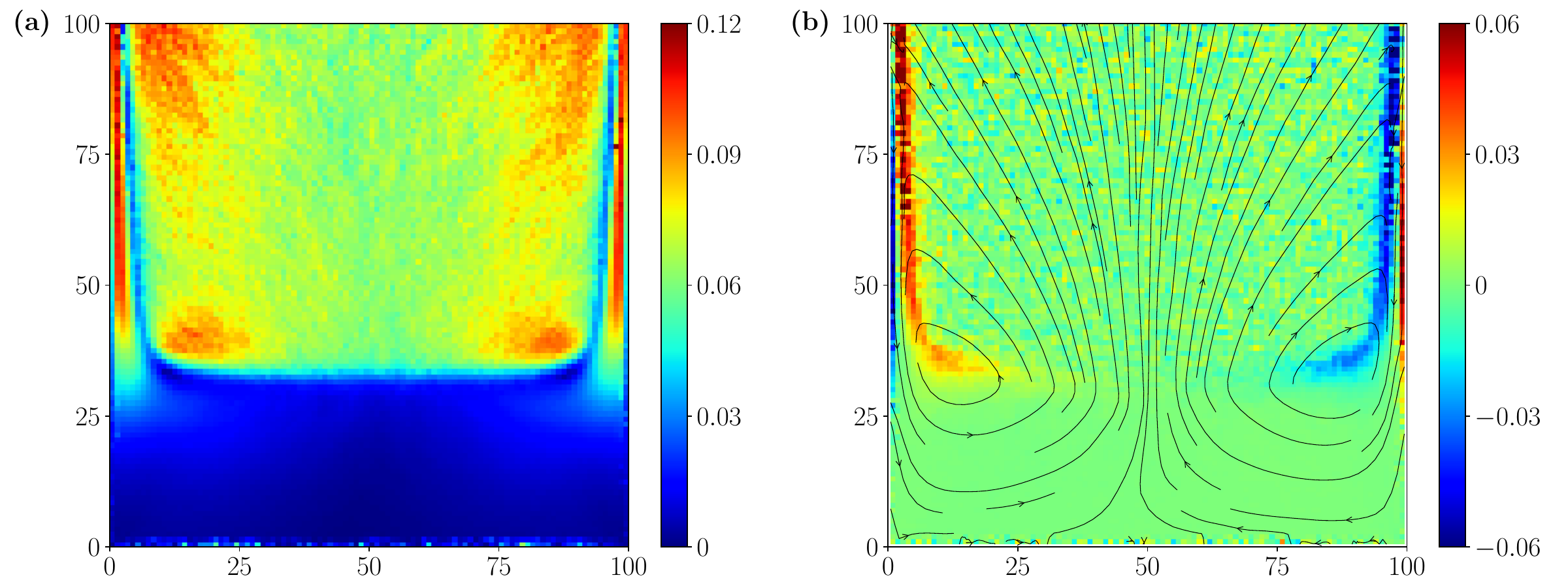}
\caption{Stationary state of ABPs in a box with reflecting walls. Box dimension is $100 \times 400$, particle number is $5000$, gravity is in $-\hat{y}$ direction, $F_0=100$, ${\rm Pe}_s=30$, and ${\rm Pe}_g=6$. Shown quantities are time-averaged. Here we consider the averaged velocity defined as ${\bf V} = {\bf J}/\rho$. (a)~Modulus of the velocity $|{\bf V}(x,y)|$. (b)~Curl amplitude $\partial_x V_y - \partial_y V_x$ together with arrows indicating velocity orientation $\phi_V(x,y)$, such that ${\bf V} = (\cos \phi_V, \sin \phi_V)$.}
\end{center}
\end{figure}

\begin{figure}[H]
\begin{center}
\includegraphics[width=12cm]{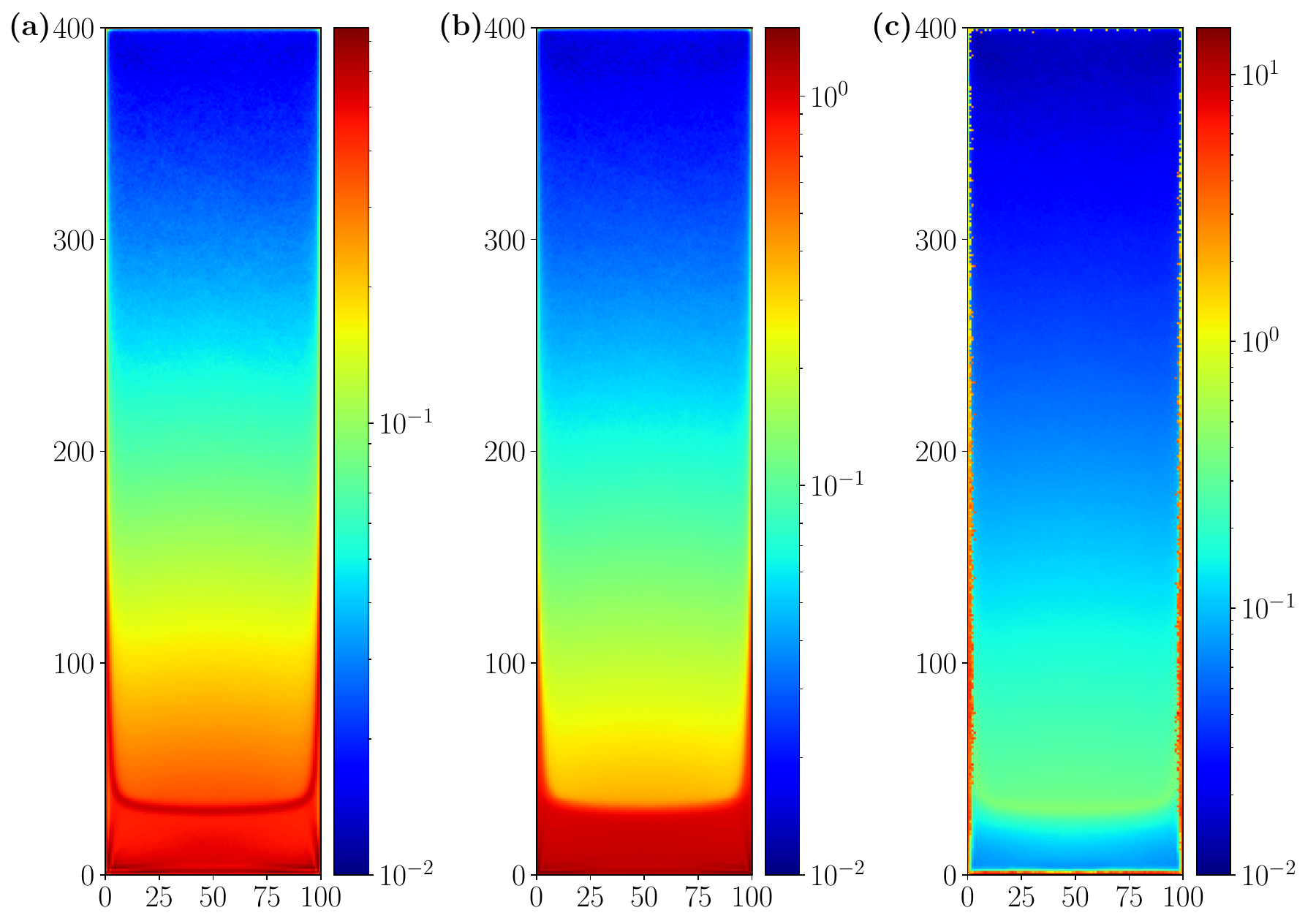}
\caption{Temporal fluctuations of the stationary state quantities shown in Fig. 1: (a)~Standard deviation of the particle density $\sigma_\rho = \sqrt{\langle \rho^2 \rangle_t - \langle \rho \rangle_t^2}$, where $\langle\cdot\rangle_t$ denotes the average over time. (b)~Standard deviation of the polarization $\sigma_P = \sqrt{\langle {\bf P}^2 \rangle_t - \langle {\bf P} \rangle_t^2}$. (c)~Standard deviation of the current density $\sigma_J = \sqrt{\langle {\bf J}^2 \rangle_t - \langle {\bf J} \rangle_t^2}$. }
\end{center}
\end{figure}

\end{document}